\documentclass[12pt,a4paper]{article}
\usepackage{latexsym}
\usepackage{amssymb}
\usepackage{amsmath}
\usepackage{amsfonts}

\numberwithin{equation}{section}

\def\a{{\alpha}}

\def\b{{\beta}}
\def\g{{\gamma}}

\def\d{{\delta}}
\def\r{{\rho}}

\def\ep{{\varepsilon}}
\def\en{{\epsilon}}

\def\z{{\zeta}}
\def\f{\frac}

\def\la{\label}
\def\eq{\eqref}

\def\nb{\nabla}
\newcommand{\be}{\begin{equation}}
\newcommand{\ee}{\end{equation}}

\newcommand{\Tr}{{\rm Tr}}

\def\cN{{\cal N}}
\def\bea{\begin{eqnarray}}
\def\eea{\end{eqnarray}}
\def\nn{\nonumber}

\begin{document}

\begin{titlepage}


\begin{center}
{\Large \bf
Two-loop low-energy effective actions in \\[3mm]
$\cN=2$ and $\cN=4$ three-dimensional SQED}

\vspace{1cm}

 {\bf
 I.L. Buchbinder $^+$\footnote{joseph@tspu.edu.ru},
 B.S. Merzlikin $^{+\dag}$\footnote{merzlikin@tspu.edu.ru},
 I.B. Samsonov $^{*}$\footnote{samsonov@mph.phtd.tpu.ru, on leave from
Tomsk Polytechnic University, 634050 Tomsk, Russia.}}
\\[3mm]
 {\it $^+$ Department of Theoretical Physics, Tomsk State Pedagogical
 University,\\ Tomsk 634061, Russia\\[2mm]
 $^\dag$
 Department of Higher Mathematics and Mathematical Physics,\\
\it Tomsk Polytechnic University, 634050 Tomsk, Russia
 \\[2mm]
 $^*$ INFN, Sezione di Padova, via F. Marzolo 8, 35131 Padova, Italy}
\end{center}

\vspace{1cm}

\begin{abstract}
We study two-loop Euler-Heisenberg effective actions in
three-dimensional $\cN=2$ and $\cN=4$ supersymmetric quantum
electrodynamics (SQED) without Chern-Simons term. We find exact
expressions for propagators of chiral superfields interacting with
slowly-varying $\cN=2$ gauge superfield. Using these propagators
we compute two-loop effective actions in the $\cN=2$ and $\cN=4$
SQED as the functionals of superfield strengths and their
covariant spinor derivatives. The obtained effective actions
contain new terms having no four-dimensional analogs. As an
application, we find two-loop quantum corrections to the moduli
space metric in the $\cN=2$ SQED.
\end{abstract}

\end{titlepage}

\setcounter{footnote}{0}

\section{Introduction}
Low-energy dynamics of three-dimensional supersymmetric gauge
theories has attracted considerable attention recently (see, e.g.,
\cite{IS13,2} and references therein). $\cN=2$ and $\cN=4$
supersymmetric Yang-Mills-Chern-Simons theories possess many
remarkable properties in classical and quantum domains such as the
mirror symmetry
\cite{IS96,deBoer1996,deBoer1996ck,deBoer97,deBoer,AHISS} and
Seiberg-like dualities \cite{deBoer,AHISS,Aharony97,Karch,GK08}. A
lot of information about low-energy dynamics is encoded in the
structure of the moduli space which comprises both perturbative
and non-perturbative effects. The perturbative quantum
contributions to such moduli spaces are known only up to one-loop
order \cite{deBoer,SW96} while the higher-loop corrections are
also of interest and deserve detailed investigations. This
motivates study of higher-loop quantum corrections to the
low-energy effective actions in three-dimensional supersymmetric
gauge theories.

In this paper, we compute two-loop Euler-Heisenberg effective
actions in $\cN=2$ and $\cN=4$ supersymmetric electrodynamics with
vanishing Chern-Simons term. The classical actions of these models
arise as a result of dimensional reduction from the
four-dimensional $\cN=1$ and $\cN=2$ SQED, respectively. The
two-loop Euler-Heisenberg effective actions in the
four-dimensional supersymmetric models were derived in
\cite{Kuz0310,Kuz07} using the technique of covariant perturbative
multiloop computations in the $\cN=1$, $d=4$ superspace \cite{Kuz03}.
The attractive features of this method are its
universality, generality and a possibility to preserve manifestly
the $\cN=1$, $d=4$ supersymmetry and gauge invariance on all
stages of loop calculations. In the present paper we extend this
technique to the three-dimensional gauge theories in the $\cN=2$,
$d=3$ superspace. In particular, we derive exact propagators of
chiral superfields on slowly-varying gauge superfield background
and apply them for computing two-loop low-energy effective actions
in the $\cN=2$ and $\cN=4$ SQED. As we show in the present paper,
the obtained effective actions possess new terms having no
four-dimensional analogs, but playing important role in the
low-energy dynamics of these models.

In general, the Euler-Heisenberg superfield effective actions in
three-dimensional $\cN=2$ and $\cN=4$ SQED can be represented in the
following general form:
\be
\Gamma=\int d^3x d^4\theta\,{\cal
L}_{\rm eff}(G,\Phi\bar\Phi,W_\alpha, \bar
W_\alpha,D_{(\alpha}W_{\beta)})\,.
\label{1}
\ee
Here ${\cal L}_{\rm eff}$ is the effective Lagrangian which
depends on the $\cN=2$ superfield strengths $G$, $W_\alpha$ and $\bar
W_\alpha$ and the chiral superfield $\Phi$ which is a part of the $\cN=4$
gauge multiplet. The $\cN=2$ case corresponds to freezing the
chiral superfield $\Phi$ to be equal to the $\cN=2$ complex mass
parameter, $\Phi=m$. The superfield strengths are assumed to be
slowly-varying such that we omit all their space-time derivatives
and keep only the dependence on
$N_{\alpha\beta}\equiv D_{(\alpha}W_{\beta)}$.\footnote{In principle,
one can consider also $\bar N_{\alpha\beta}\equiv \bar D_{(\alpha}\bar
W_{\beta)}$,
but unlike the four-dimensional case this expression in not independent,
$\bar N_{\alpha\beta}=-N_{\alpha\beta}$.}
 In components, the action of the form (\ref1)
contains all powers of the Maxwell field strength $F^{2n}$ with
their supersymmetric completions. One-loop Euler-Heisenberg
effective actions in the $\cN=2$ and $\cN=4$ SQED were computed in
\cite{BPS1}.

The part of the effective Lagrangian in (\ref1) which depents only
on $G$ and $\Phi\bar\Phi$ we refer to as the effective potential. In
the $\cN=4$ gauge theory, the one-loop effective potential was
derived more than a quarter of a century ago in
\cite{HKLR} from geometrical principles,\footnote{This is a
three-dimensional analog of the $\cN=2$, $d=4$ improved tensor
multiplet superspace action \cite{LR}.
See also \cite{book}.}
\be f(G,\Phi\bar\Phi)={\cal L}_{\rm eff}|_{W_\alpha=\bar
W_\alpha=0} \propto G \ln (G+\sqrt{G^2+\bar\Phi\Phi})
-\sqrt{G^2+\Phi\bar\Phi}\,. \label{2}
\ee
It is the effective
potential which is responsible for the moduli space metric
\cite{HKLR,deBoer}. Note that in the $\cN=4$ gauge theory the
effective potential (\ref2) is one-loop exact, but in the $\cN=2$
theories it can receive higher-loop quantum contributions. In the
present paper we compute two-loop effective superpotential in the
$\cN=2$ SQED and find corresponding two-loop quantum corrections to
the moduli space metric. To the best of our knowledge, these two-loop
corrections to the moduli space metric have not been presented
before.

The rest of the paper is organized as follows. In Sect.\ 2 we study
the basic properties of parallel displacement propagator in $\cN=2$,
$d=3$ superspace and derive exact propagators of real and chiral
superfields in slowly-varying gauge superfield background. In
Sect.\ 3 we employ these propagators for computing the two-loop
low-energy Euler-Heisenberg effective action in the $\cN=2$
supersymmetric electrodynamics. As an application of the obtained
effective action, we find two-loop quantum corrections to the moduli
space metric in the $\cN=2$ supergauge theory. Section 4 is devoted
to computing the effective action in the $\cN=4$ SQED. In the last
section we summarize the obtained results and discuss their possible
generalizations. In Appendices we collect $\cN=2$ superspace
notations exploited throughout the text and give some technical
details of computations.


\section{Exact propagators on gauge superfield background}
\subsection{Gauge theory in $\cN=2$, $d=3$ superspace}
The $\cN=2$, $d=3$ superspace is parametrized by the coordinates
$z^A=(x^m,\theta^\alpha,\bar\theta_\alpha)$, $m=0,1,2$. The
corresponding supercovariant derivatives $D_A=(\partial_m,D_\alpha,\bar
D^\alpha)$ are written down in (\ref{Dexpl}).

The (Abelian) gauge superfields in the $\cN=2$ superspace can be
introduced within the standard geometric approach based on adding
gauge connections $V_A=(V_m,V_\alpha,\bar V^\alpha)$ to the
``flat'' superspace derivatives,
\be D_\alpha\to \nabla_\alpha=
D_\alpha+V_\alpha\,,\quad \bar D_\alpha\to \bar\nabla_\alpha=\bar
D_\alpha+\bar V_\alpha\,,\quad
\partial_m \to\nabla_m =\partial_m + V_m\,,
\label{2.2}
\ee
and imposing the superfield constraints \cite{HKLR,ZP,NG},
\bea
\{\nabla_\alpha,\bar\nabla_\beta \}&=&-2i(\gamma^m)_{\alpha\beta}
\nabla_m +2i\varepsilon_{\alpha\beta}G\,,
\nn
\\
{}[\nabla_\alpha,\nabla_m]&=&-(\gamma_m)_{\alpha\beta}\bar
W^\beta\,,\qquad
[\bar\nabla_\alpha,\nabla_m]=(\gamma_m)_{\alpha\beta}
W^\beta\,,
\nn
\\
{}[\nabla_m,\nabla_n]&=&iF_{mn}\,.
\label{gauge-alg}
\eea
The superfield strengths $G$, $W_\alpha$ and $\bar W_\alpha$
in this algebra satisfy the following reality
properties:
\be
G^*=G\,,\quad
(W^\alpha)^*=\bar W^\alpha\,,\quad
(F_{mn})^*=F_{mn}\,.
\ee

The algebra (\ref{gauge-alg}) possesses many Bianchi identities. In
particular, the superfield strength $G$ is linear,
\be
D^2 G= \bar D^2 G=0\,, \label{linearity}
\ee
while $W_\alpha$ and $\bar
W_\alpha$ can be expressed in terms of $G$,
\be
W_\alpha = \bar
D_\alpha G\,,\qquad \bar W_\alpha = D_\alpha G\,.
\ee
As a consequence, $W_\alpha$ and $\bar W_\alpha$ are (anti)chiral,
\be
\bar D_\alpha W_\beta=0\,,\qquad D_\alpha \bar W_\beta=0\,,
\ee
and obey the `standard' Bianchi identity,
\be
D^\alpha W_\alpha = \bar D^\alpha \bar W_\alpha\,.
\ee
The superfield strength $F_{mn}$ is
also non-independent since it can be expressed in terms of other
superfields,
\be
F_{mn}=\frac14\varepsilon_{mnp}(\gamma^p)^{\alpha\beta}(D_\alpha
W_\beta
 - \bar D_\alpha \bar W_\beta)\,.
\ee
Finally, there is one more useful identity which involves
space-time derivative of $G$,
\be
\partial_m G= \frac i4 \gamma_m^{\alpha\beta}
(D_\alpha W_\beta + \bar D_\alpha \bar W_\beta)\,.
\label{2.9}
\ee

The algebra (\ref{gauge-alg}) is invariant under the following gauge
transformations,
\be \nabla_A \to e^{i\tau(z)} \nabla_A
e^{-i\tau(z)}\,,\qquad \tau^* = \tau\,,
\label{gauge-tau}
\ee
with $\tau(z)$ being arbitrary real gauge parameter.

Let us introduce a real gauge superfield, $V=V^*$, and represent
the gauge connections $V_A$ in terms of it,
\be
\nabla_\alpha=e^{-2V} D_\alpha e^{2V} = D_\alpha + 2 D_\alpha
V\,,\qquad \bar\nabla_\alpha = \bar D_\alpha\,.
\ee
The algebra (\ref{gauge-alg}) leads to the following expressions
for the superfield strengths,
\be
G=\frac i2\bar D^\alpha D_\alpha V\,,\quad
W_\alpha = -\frac i4 \bar D^2 D_\alpha V\,,\quad
\bar W_\alpha = -\frac i4 D^2 \bar D_\alpha V\,.
\ee

For the problem of Euler-Heisenberg effective action it is
sufficient to consider the background gauge superfield which obeys
the following constraints:
\begin{itemize}
\item[(i)] $\cN=2$ supersymmetric Maxwell equations,
\be
D^\alpha W_\alpha = 0\,,\qquad
\bar D^\alpha \bar W_\alpha =0\,;
\label{constr1}
\ee
\item[(ii)]
The superfield strengths are constant with respect to the
space-time derivative,
\be
\partial_m G=0\,,\quad
\partial_m W_\alpha =0 \,,\quad
\partial_m \bar W_\alpha =0 \,.
\label{constr2}
\ee
\end{itemize}
The latter constraint means that we consider a slowly-varying
gauge superfield background.

\subsection{Parallel displacement propagator in ${\cal N}=2$, $d=3$ superspace}

It is well known that quantization of gauge theories requires
gauge fixing and, as a consequence, all off-shell quantities in
gauge theories are gauge dependent. As to the effective
action, it can be formulated in such a way that being gauge
dependent it remains invariant under the classical gauge
transformations. This formulation is called the background field
method. The main idea of this method is a splitting of the gauge
field into `background' and `quantum' parts and imposing the gauge
fixing only on the quantum field. Such a gauge fixing condition is
taken to be background field dependent that provides classical
gauge invariance of the effective action.

Quantum loop calculations within the background field method assume to
operate with the background field dependent propagators which, in
general, cannot be written in an explicit form. For the problem of
low-energy effective action, it is sufficient to represent these propagator as
series in power of field strengths and their covariant derivatives. Such
propagators are naturally obtained on the basis of proper-time
technique which allows one to develop manifestly gauge invariant
procedure for computing the one-loop effective action. Superfield
proper-time technique and its application for finding the superfield
effective actions is described e.g.\ in \cite{book}. However,
manifestly gauge invariant computations of multiloop contributions
to effective actions require new methods in
comparison with the one-loop computations. One of such efficient methods
is based on the employment of the parallel displacement
propagator\footnote{The use of parallel displacement propagator for
quantum field theory in curved space-time was initiated by DeWitt
\cite{DeWitt}.}.

The technique of multiloop quantum computations in the $\cN=1$,
$d=4$ superspace which involves the parallel displacement
propagator was elaborated in \cite{Kuz03}. The power of this
method was demonstrated, in particular, in the studies of two-loop
effective actions in the four-dimensional $\cN=1$ and $\cN=2$
SQED \cite{Kuz0310,Kuz07}. Our aim is to extend this
technique for the $\cN=2$, $d=3$ superfield gauge theories. In this
section we study basic properties of the parallel
displacement propagator associated with the algebra
(\ref{gauge-alg}). The obtained formulae will be applied in the
next section for two-loop quantum computations of low-energy
effective actions in the three-dimensional $\cN=2$ and $\cN=4$
supersymmetric electrodynamics.

By definition, the parallel displacement propagator $I(z,z')$ is a two-point
superspace function depending on the gauge superfields
with the following properties:
\begin{itemize}
\item[(i)]
Under the gauge transformations (\ref{gauge-tau}) it transforms as
\be
I(z,z')\to e^{i\tau(z)}I(z,z')e^{-i\tau(z')}\,;
\label{prop1}
\ee
\item[(ii)]
It obeys the equation
\be
\zeta^A \nabla_A I(z,z')=\zeta^A\left(D_A +
V_A(z)\right)I(z,z')=0\,,
\label{prop2}
\ee
where $\zeta^A=(\rho^m,\zeta^\alpha,\bar\zeta_\alpha)$ is the
$\cN=2$ supersymmetric interval,
\be
\zeta^\alpha=(\theta-\theta')^\alpha\,,\quad
\bar\zeta^\alpha= (\bar\theta-\bar\theta')^\alpha\,,\quad
\rho^m=(x-x')^m-i\gamma^m_{\alpha\beta}\zeta^\alpha \bar\theta'^\beta
+i\gamma^m_{\alpha\beta}\theta'^\alpha \bar\zeta^\beta\,;
\label{superinterval}
\ee
\item[(iii)] For coincident superspace points $z=z'$ it reduces to the
identity operator in the gauge group,
\be
I(z,z)=1\,.
\label{prop3}
\ee
\end{itemize}

One can show that the properties (\ref{prop1}) and (\ref{prop3})
imply the important identity
\be
I(z,z')I(z',z)=1\,.
\label{id-unit}
\ee

The rule of Hermitian conjugation for $I(z,z')$ looks like
 \be
\left( I(z,z') \right)^\dag = I(z',z)\,. \ee

It is convenient to rewrite the algebra of gauge-covariant derivatives
(\ref{gauge-alg})  in the following condensed form,
\bea
 [\nb_A,\nb_B\}={\bf T}_{A\,B}{}^{C}\nb_C+i{\bf F}_{A\,B}\,,
 \la{supal}
\eea
where ${\bf T}_{A\,B}{}^{C}$ is a supertorsion and ${\bf
F}_{A\,B}$ is a supercurvature for gauge superfield connections
(\ref{2.2}). In \cite{Kuz03} it was proved that, owing to
(\ref{prop2}), the action
of the derivative $\nabla_B$ on $I(z,z')$ can be expressed in terms
${\bf T}_{AB}{}^C$, ${\bf F}_{AB}$ and their covariant
derivatives,
 \bea
\nb_B I(z,z')&=&i I(z,z')\sum_{n=1}^\infty \f{1}{(n+1)!}\,\bigg[n\zeta^{A_n}
 \ldots\zeta^{A_1}\nb'_{A_1}\ldots\nb'_{A_{n-1}}{\bf F}_{A_n\,B}(z')  \nn\\
&&+ \f{(n-1)}{2}\zeta^{A_n}{\bf T}_{A_n\,B}{}^{C} \zeta^{A_{n-1}}
\ldots\zeta^{A_1}\nb'_{A_1}\ldots\nb'_{A_{n-2}}{\bf F}_{A_{n-1}\,C}(z')
 \bigg]\,,\la{dec1}
 \eea
There is also an equivalent form of this relation in which
$I(z,z')$ appears on the right,
\bea
\la{dec2}
\nb_B I(z,z')&=&i \sum_{n=1}^\infty \f{(-1)^n}{(n+1)!}\,\bigg[-\zeta^{A_n}
 \ldots\zeta^{A_1}\nb_{A_1}\ldots\nb_{A_{n-1}}{\bf F}_{A_n\,B}(z)\\
&& + \f{(n-1)}{2}\zeta^{A_n}{\bf T}_{A_n\,B}{}^{C} \zeta^{A_{n-1}}
 \ldots\zeta^{A_1}\nb_{A_1}\ldots\nb_{A_{n-2}}{\bf F}_{A_{n-1}\,C}(z)
 \bigg]I(z,z')\,.\nn
 \eea

Recall that we consider the gauge superfield background which obeys
the constraints (\ref{constr1}) and (\ref{constr2}). For such a
background the serieses in (\ref{dec1}) and (\ref{dec2}) terminate
and we obtain:
\bea \label{DI1} \nabla_\beta I(z,z')&=&\bigg[
-i\bar\zeta_\beta G +\frac12\r_{\alpha\beta} \bar W^\alpha-\frac
i{12}\bar\zeta^2 W_\beta +\frac i6\bar\zeta_\beta\zeta^\alpha \bar
W_\alpha -\frac i3 \bar\zeta^\alpha \zeta_\alpha \bar W_\beta
\\
&&+\frac1{12} \bar \zeta^\alpha \r_{\beta\gamma}
\bar \nb_\alpha \bar W^\gamma
-\frac1{12} \bar\zeta^\alpha \r_{\alpha\gamma}
\bar\nb^\gamma \bar W_\beta
-\frac i{12}\bar\zeta^2 \zeta_\beta \bar\nb^\alpha \bar W_\alpha
\bigg]I(z,z')\nn\\
&=&I(z,z')\bigg[
-i\bar\zeta_\beta G +\frac12\r_{\alpha\beta} \bar W^\a
-\frac{7 i}{12}\bar\zeta^2 W_\beta
-\frac{5i}6\bar\zeta_\beta\zeta^\alpha \bar W_\alpha
-\frac i3 \bar\zeta^\alpha \zeta_\alpha \bar W_\beta
\nn\\&&
-\frac5{12}\bar \zeta^\alpha \r_{\beta\gamma}
\bar\nb_\alpha \bar W^\gamma
-\frac1{12} \bar\zeta^\alpha \r_{\a\g} \bar\nb^\gamma \bar W_\beta
+\frac i3\bar \zeta^2 \zeta^\alpha \bar\nb_\beta \bar W_\alpha
+\frac i{12}\bar \zeta^2 \zeta_\beta \bar\nb^\alpha \bar W_\alpha
\bigg]\,, \nn\\
\label{DI2}
\bar\nabla^\beta I(z,z')&=&\bigg[
-i \zeta^\beta G - \frac 12 \r_{\alpha}^\beta W^\alpha
+\frac i{12} \zeta^2 \bar W^\beta
- \frac i6 \zeta^\beta \bar\zeta^\alpha W_\alpha
+\frac i3 \zeta^\alpha \bar \zeta_\alpha W^\beta
\\&&
+\frac1{12}\zeta_\alpha \r^{\beta\gamma}\nb^\alpha W_\gamma
-\frac1{12}\z_\a  \r^{\alpha\gamma} \nb_\gamma W^\beta
-\frac i{12}\z^2 \bar\z^\b \nb^\a W_\a\bigg]I(z,z')
\nn\\&
=&I(z,z')\bigg[
-i\z^\b G -\f12\r_{\a}^\b W^\a
+\frac{7i}{12}\z^2 \bar W^\b + \f{5i}6 \z^\b\bar\z^\a W_\a
+\frac i3\zeta^\alpha \bar\zeta_\alpha W^\beta
\nn\\&&
+\frac5{12} \z^\a\r^{\beta\gamma} \nb_\a W_\g
-\frac1{12} \z_\a \r^{\a\g}\nb_\g W^\b
-\frac i3 \z^2 \bar \z_\a \nb^\b W^\a
+\frac i{12} \z^2\bar\z^\b \nb^\a W_\a\bigg], \nn\\
\nabla_m I(z,z')&=&\bigg[
\frac i2\r^n F_{nm}
-\frac12(\gamma_m)_{\alpha\beta}\Big(\zeta^\alpha \bar W^\beta
+\bar\zeta^\alpha W^\beta
\nn\\&&
+\frac13\zeta^\alpha\bar\zeta^\gamma \bar\nb_\gamma \bar W^\beta
-\frac13\bar\zeta^\alpha\zeta^\gamma \nb_\gamma W^\beta \Big)
\bigg]I(z,z')\nn\\
&=&I(z,z')\bigg[
\frac i2\r^n F_{nm}
-(\gamma_m)_{\alpha\beta}\Big(\f12\zeta^\alpha \bar W^\beta
+\f12\bar\zeta^\alpha W^\beta
\nn\\&&
-\frac13 \zeta^\alpha\bar\zeta^\gamma \bar\nb_\gamma \bar W^\beta
+\frac13\bar\zeta^\alpha\zeta^\gamma \nb_\gamma W^\beta\Big)
\bigg].
\label{DI3}
\eea

In comparison with the four-dimensional case, the expressions
(\ref{DI1},\ref{DI2},\ref{DI3}) involve the superfield $G$ which
will lead to new contributions in the effective action having
no four-dimensional analogs.


\subsection{Real superfield Green's function and its heat kernel}

There are three basic d'Alembertian-like operators which occur in
covariant supergraphs \cite{BPS1,BPS2}: (i) the d'Alembertian $\square_{\rm
v}$ which acts in the space of real superfields;
(ii) the chiral d'Alembertian $\square_+$ acting on chiral superfields; and (iii) the
antichiral d'Alembertian $\square_-$. The latter is related to the
former by conjugation. Therefore we concentrate mainly on $\square_{\rm
v}$ and $\square_+$.

The real superfield d'Alembertian is defined by two equivalent lines:
\bea
\square_{\rm v}
&=&
-\frac18\nabla^\alpha\bar\nabla^2 \nabla_\alpha
 +\frac1{16}\{\nabla^2,\bar\nabla^2 \} + \frac i2(\nabla^\alpha W_\alpha)
 +iW^\alpha \nabla_\alpha
\label{first}
\nn \\
&=&-\frac18 \bar\nabla^\alpha \nabla^2 \bar\nabla_\alpha
 +\frac1{16}\{\nabla^2 ,\bar\nabla^2  \} - \frac i2
 (\bar\nabla^\alpha \bar W_\alpha) -i \bar W^\alpha
 \bar\nabla_\alpha\,.
\label{second} \eea By virtue of the algebra (\ref{gauge-alg}) it
can be represented in the following form,
\be \square_{\rm v}=
\nabla^m \nabla_m+G^2 +iW^\alpha \nabla_\alpha -i \bar W^\alpha
\bar\nabla_\alpha\,. \ee
Let us consider Green's function for this
operator $G_{\rm v}(z,z')$ and the corresponding heat kernel $K_{\rm
v}(z,z'|s)$, \be (\square_{\rm v}+m^2)G_{\rm
v}(z,z')=-\delta^7(z-z')\,,\qquad G_{\rm v}(z,z')=i\int_0^\infty
ds\, K_{\rm v}(z,z'|s)e^{is(m^2+i\epsilon)}\,, \ee
where $m$ is a
mass parameter and $\epsilon\to+0$ implements standard
boundary condition for the propagator. For the  gauge superfield
background (\ref{constr1},\ref{constr2}) the explicit expression for
$K_{\rm v}$ was derived in \cite{BPS1}, \bea K_{\rm
v}(z,z'|s)=\frac{1}{8(i\pi s)^{3/2}}\frac{sB}{\sinh(sB)} {\cal
O}(s)\,e^{isG^2}\,
e^{\f{i}{4}(F\coth(sF))_{mn}\r^n\r^m}\z^2\bar\z^2\,I(z,z')\,,
\label{ker2} \eea where $\zeta^2=\zeta^\alpha\zeta_\alpha$,
$\bar\zeta^2 = \bar\zeta^\alpha \bar\zeta_\alpha$ and $\rho^m$ are
the components of the supersymmetric interval (\ref{superinterval})
and the following notations are employed, \be {\cal O}(s)=e^{s(\bar
W^\a\bar \nb_\a-W^\a\nb_\a)}\,,\la{O} \ee \be B^2 =
\frac12N_\alpha^\beta N_\beta^\alpha\,,\quad
N_{\alpha\beta}=D_{(\alpha} W_{\beta)}\,,\quad \bar
N_{\alpha\beta}=\bar D_{(\alpha} \bar W_{\beta)}\,. \label{B2} \ee
Note that the parallel displacement propagator $I(z,z')$ in
(\ref{ker2}) provides the correct transformation properties of the
heat kernel under the gauge symmetry (\ref{gauge-tau}). Note also
that, owing to (\ref{2.9}), for the case of constant superfield
background (\ref{constr2}) $\bar N_{\alpha\beta}$ is not
independent, but coincides with $N_{\alpha\beta}$ up to sign,
$\bar N_{\alpha\beta} =-N_{\alpha\beta}$. Therefore we will use only
$N_{\alpha\beta}$ in what follows.

The expression (\ref{ker2}) contains the operator ${\cal O}(s)$
which acts both on the superfields and on the components of
the supersymmetric interval. Let us push this operator on the
right and act with it on the parallel displacement propagator,
\bea
K_{\rm v}(z,z'|s)=\f{1}{8(i\pi s)^{3/2}}\f{sB}{\sinh(sB)}
e^{isG^2}\,e^{\f{i}{4}(F\coth(sF))_{mn}\r^n(s)\r^m(s)}\z^2(s)\bar\z^2(s)I(z,z'|s)\,,
\la{ker3}
\eea
where the following notations have been introduced:
\bea
W^\alpha(s)&\equiv&{\cal O}(s)W^\alpha {\cal O}(-s)
=W^\beta (e^{-sN})_\beta{}^\alpha\,,\nn\\
\zeta^\alpha(s)&\equiv&{\cal O}(s)\zeta^\alpha {\cal O}(-s)
= \zeta^\alpha+W^\beta
((e^{-sN}-1)N^{-1})_\beta{}^\alpha\,,\nn\\
\bar\zeta^\alpha(s)&\equiv&{\cal O}(s)\bar\zeta^\alpha {\cal O}(-s)
= \bar\zeta^\alpha+\bar W^\beta
((e^{-s N}-1)N^{-1})_\beta{}^\alpha\,,\nn\\
\rho^m(s)&\equiv&{\cal O}(s)\rho^m{\cal O}(-s)
=\rho^m-i(\gamma^m)^{\alpha\beta}\int_0^s dt\left(W_{\alpha}(t)\bar\zeta_{\beta}(t)
+\bar W_{\alpha}(t)\zeta_{\beta}(t)\right)\,,\la{id's}
\eea
and
\be
I(z,z'|s)\equiv {\cal O}(s)I(z,z')\,.
\ee

Owing to (\ref{DI1}) and (\ref{DI2}),
the expression for $I(z,z'|s)$ can be written explicitly in terms
of superfield strengths and their derivatives. Indeed, by
differentiating over the proper time $s$, it is easy to check the identity
\be
I(z,z'|s)= \exp\left[\int_0^s dt\, \Sigma(z,z'|t) \right] I(z,z')\,,
\label{I(s)}
\ee
where
\be
\Sigma(z,z'|t)={\cal O}(t) \Sigma(z,z'){\cal O}(-t)\,,
\label{Sigma(t)}
\ee
and $\Sigma(z,z')$ solves
\be
(\bar W^\alpha \bar\nabla_\alpha - W^\alpha \nabla_\alpha)I(z,z')
=\Sigma(z,z') I(z,z')\,.
\ee
Applying \eq{DI1} and \eq{DI2} in the latter equation we
immediately find $\Sigma(z,z')$,
 \bea
\Sigma(z,z')&=&-i(\bar W^\beta \zeta_\beta - W^\beta \bar \zeta_\beta)G
-\frac i3 \zeta^\alpha \bar \zeta^\beta W_\beta \bar W_\alpha
+\frac {2i}3 \zeta^\alpha \bar\zeta_\alpha W^\beta \bar W_\beta
\nn\\&&
+\frac i{12}\zeta^2 [\bar W^2-\bar\zeta^\alpha\bar W_\alpha D^\beta W_\beta]
+\frac i{12}\bar\zeta^2[W^2+\zeta^\alpha W_\alpha \bar D^\beta \bar W_\beta]
\nn\\&&
+\frac1{12}(\zeta^\alpha \bar W^\beta -\bar\zeta^\beta W^\alpha)
[\r_{\alpha\gamma} D^\gamma W_\beta+\r_{\beta\gamma}
 \bar D^\gamma \bar W_{\alpha}]\,.
 \la{Sigma}
 \eea
The expression for $\Sigma(z,z'|s)$ appears from $\Sigma(z,z')$ by simply
making all ingredients of (\ref{Sigma}) $s$-dependent as in
(\ref{id's}).

\subsection{Chiral Green's function and its heat kernel}

The d'Alembertian operators acting in the space of (anti)chiral
superfields read \cite{BPS1,BPS2}
\be
\square_+=\nabla^m\nb_m+G^2+\f{i}{2}(\nb^\a W_\a) +i W^\a\nb_\a \,,
\quad
\square_+\Phi=\frac1{16}\bar\nabla^2 \nabla^2 \Phi\,,\quad
\bar\nabla_\alpha\Phi=0\,,
 \la{D_+}
\ee
\be
\square_-=\nabla^m \nabla_m +G^2 - \frac i2 (\bar\nabla^\alpha \bar W_\alpha)
-i\bar W^\alpha \bar \nabla_\alpha\,,\quad
\square_-\bar\Phi=\frac1{16}\nabla^2 \bar\nabla^2\bar\Phi\,,\quad
\nabla_\alpha\bar\Phi=0\,.
\label{D_-}
\ee
The Green's functions for these operators obey
\be
(\square_++m^2)G_+(z,z')=-\delta_+(z,z')\,,\qquad
(\square_-+m^2)G_-(z,z')=-\delta_-(z,z')\,,
\label{2.43}
\ee
where $\delta_\pm(z,z')$ are (anti)chiral delta-functions. These
Green's functions are expressed in terms of the corresponding heat
kernels,
\be
G_\pm(z,z')=i\int_0^\infty ds\,
K_\pm(z,z'|s)e^{is(m^2+i\epsilon)}\,,\qquad
\epsilon\to+0\,.
\ee

The operators (\ref{D_+}) and (\ref{D_-}) are related to each
other as
\be
\nabla^2\square_+=\square_-\nabla^2\,, \qquad
\bar\nabla^2\square_-=\square_+\bar\nabla^2\,.
 \la{id1}
 \ee
Moreover, when the background gauge superfield obeys
supersymmetric Maxwell equations (\ref{constr1}),
these operators are related to $\square_{\rm v}$,
\be
\nabla^2\square_+=\nabla^2\square_{\rm v}=\square_{\rm v}\nabla^2 \,, \qquad
\bar\nabla^2\square_-=\bar\nabla^2\square_{\rm v}=\square_{\rm v}\bar\nabla^2\,.
\la{id2}
 \ee
As a consequence of these identities, the (anti)chiral
Green's functions can be expressed in terms of $G_{\rm v}$,
\be
G_+(z,z')=-\frac14\bar\nabla^2 G_{\rm v}(z,z')\,,\qquad
G_-(z,z')=-\frac14 \nabla^2 G_{\rm v}(z,z')\,,
\ee
and similar relations hold for the corresponding heat kernels,
\be
K_+(z,z'|s)=-\frac14\bar\nabla^2 K_{\rm v}(z,z'|s)\,, \qquad
K_-(z,z'|s)=-\frac14\nabla^2 K_{\rm v}(z,z'|s)\,. \la{ker4}
\ee

To compute $K_+$ we have to differentiate (\ref{ker2}) by
$\bar\nabla^2$. Owing to the
identities (\ref{id2}), the operator $\bar\nabla^2$ acts only on
$\bar\zeta^2 I(z,z')$,
\be
K_+(z,z'|s)=\f{1}{8(i\pi s)^{3/2}}\f{sB}{\sinh(sB)}
{\cal O}(s)\,e^{isG^2}\,
e^{\f{i}{4}(F\coth(sF))_{mn}\r^n\r^m}\z^2
\left(-\f14\bar\nb^2\right) \bar\zeta^2\,I(z,z')\,.
\la{ker7}
\ee
The action of the derivative $\bar\nabla_\alpha$ on $I(z,z')$ is
given by (\ref{DI2}). However, only one term from this expression
survives owing to $\z^\a \z^\b \z^\g=0$ and we get
\be
 -\f14\zeta^2\bar \nabla^2 (\bar\zeta^2 I(z,z'))=
 \zeta^2 e^{-\frac12\bar\zeta^\alpha\r_{\alpha\beta}W^\beta}I(z,z')\,.
\la{od}
 \ee
Substituting \eq{od} into \eq{ker7} we find the chiral heat kernel
in the following form
\be
K_+(z,z'|s)=
\f{1}{8(i\pi s)^{3/2}}\f{sB}
{\sinh(sB)}e^{isG^2}{\cal O}(s)e^{\f{i}{4}(F\coth(sF))_{mn}\r^m\r^n-
\f12\bar\z^\b \r_{\b\g}W^\g} \z^2 I(z,z')\,.
\la{K_+}
\ee
Using the properties of parallel displacement propagator
(\ref{DI2}) one can check that this expression for $K_+$ is chiral
with respect to both arguments.

The formula (\ref{K_+}) contains the operator ${\cal O}(s)$ given
in (\ref{O}). Similarly as for the heat kernel $K_{\rm v}$, we
push this operator on the right,
\be
K_+(z,z'|s)=
\f{1}{8(i\pi s)^{3/2}}\f{sB}
{\sinh(sB)}e^{isG^2}e^{\f{i}{4}(F\coth(sF))_{mn}\r^m(s)\r^n(s)-
\f12\bar\z^\b(s) \r_{\b\g}(s)W^\g(s)} \z^2(s) I(z,z'|s)\,.
\label{K+fin}
\ee
All $s$-dependent objects in this expression are given explicitly
in (\ref{id's}) and (\ref{I(s)}).

The computation of the antichiral heat kernel $K_-$ goes along similar
lines with the following outcome:
\be
K_-(z,z'|s)=
\f{1}{8(i\pi s)^{3/2}}\f{sB}
{\sinh(sB)}e^{isG^2}{\cal O}(s)e^{\f{i}{4}(F\coth(sF))_{mn}\r^m\r^n-
\f12\z^\b \r_{\b\g}\bar W^\g}\bar\z^2 I(z,z')\,.
\la{K_-}
\ee

Note that the expressions for (anti)chiral heat kernels
\eq{K_+} and  \eq{K_-} are very similar to the ones in the
four-dimensional supersymmetric gauge theory given in
\cite{Kuz0310,Kuz03}.

\subsection{Green's function $G_{+-}$ and its heat kernel}

Let $\Phi$ be a covariantly chiral superfield,
$\bar\nabla_\alpha\Phi=0$.
The Green's function $G_+(z,z')$ considered in the previous
section corresponds to the propagator of the covariantly chiral superfield,
\be
i\langle \Phi(z)\Phi(z')\rangle = mG_+(z,z')\,.
\ee
It is important to consider also the chiral-antichiral propagators,
\be
i\langle \Phi(z)\bar\Phi(z')\rangle = G_{+-}(z,z')\,,\qquad
i\langle \bar\Phi(z)\Phi(z')\rangle = G_{-+}(z,z')\,.
\ee
By definition, these Green's functions obey
\bea
\frac14 \nabla^2 G_{+-}(z,z')+ m^2
G_{-}(z,z')&=&-\delta_-(z,z')\,,\nn\\
\frac14 \bar\nabla^2 G_{-+}(z,z')+ m^2
G_{+}(z,z')&=&-\delta_+(z,z')\,.
\label{G+-def}
\eea
Consider also the corresponding heat kernels,
\be
\label{G+-K}
G_{+-}(z,z')=i\int_0^\infty ds\,
K_{+-}(z,z'|s)e^{is(m^2+i\epsilon)}\,,\quad
G_{-+}(z,z')=i\int_0^\infty ds\,
K_{-+}(z,z'|s)e^{is(m^2+i\epsilon)}\,,
\ee
where the standard $\epsilon\to+0$ prescription is assumed.

Taking into account the definitions of the (anti)chiral d'Alembertians
(\ref{D_+}) and (\ref{D_-})
it is easy to see that the solutions of the equations (\ref{G+-def})
can be expressed in terms of $G_\pm$ as
\be
G_{+-}(z,z')=\frac14 \bar\nabla^2 G_-(z,z')\,,\qquad
G_{-+}(z,z')=\frac14 \nabla^2 G_+(z,z')\,,
\ee
where $G_\pm$ obey (\ref{2.43}). Similar relations hold for
the corresponding heat kernels,
\be
K_{+-}(z,z'|s)=\frac14 \bar\nabla^2 K_-(z,z'|s)\,,\qquad
K_{-+}(z,z'|s)=\frac14 \nabla^2 K_+(z,z'|s)\,,
\ee
where $K_+$ and $K_-$ are given by (\ref{K_+}) and (\ref{K_-}),
respectively.

In what follows we consider only the heat kernel $K_{+-}$. It is
obtained from $K_-$ by acting on it with the operator $\bar\nabla^2$.
Owing to the identities (\ref{id2}), this operator commutes
trivially with the expression
$e^{isG^2}{\cal O}(s)e^{\f{i}{4}(F\coth(sF))_{mn}\r^m\r^n}$ in
(\ref{K_+}) and we need only to find the action of $\bar\nabla^2$
on the rest. This procedure is quite tedious since it involves
numerous differentiation of the components of the supersymmetric
interval, superfield strengths and the parallel displacement
propagator. For the latter we have to apply the identity
(\ref{DI2}). The result can be encoded in one function $R(z,z')$
as follows
 \be
 -\f14\bar\nabla^2\left(
e^{-\frac12\zeta^\alpha\r_{\alpha\beta}\bar W^\beta}\bar\zeta^2 I(z,z')\right)
= e^{R(z,z')} I(z,z')\,,
 \ee
where
 \bea
R(z,z')&=&-i\zeta \bar \zeta G
+\frac{7 i}{12}\bar \zeta^2 \zeta W
+\frac{i}{12} \zeta^2 \bar\zeta \bar W
-\frac12  \bar\zeta^\alpha \tilde\r_{\alpha\beta} W^\beta
-\frac12  \zeta^\alpha \tilde\r_{\alpha\beta} \bar W^\beta
\nn\\&&
+\frac1{12} \zeta^\alpha\bar\zeta^\beta
 [\tilde\rho_\beta^\gamma D_\alpha  W_\gamma
 -7\tilde\rho_\alpha^\gamma  D_\gamma  W_\beta]\,.
 \la{R}
 \eea
Here $\tilde\rho_{\alpha\beta}=\gamma^m_{\alpha\beta}\tilde
\rho_m$, and $\tilde \rho_m$ is a chiral version of
the supersymmetric interval $\rho_m$,
\be
\tilde \r^{\,m}=
\r^m+i \zeta^\alpha \gamma^m_{\alpha\beta}\bar\zeta^\beta\,,
\qquad
D'_\alpha \tilde\rho^m = \bar D_\alpha \tilde\rho^m =0\,.
\la{chirho}
\ee
Given the function $R(z,z')$, we get the following expression for
the heat kernel $K_{+-}$,
\be
K_{+-}(z,z'|s)=-\frac1{8(i\pi s)^{3/2}}\frac{sB}{\sinh(sB)}
e^{isG^2}{\cal O}(s)
e^{\frac i4(F\coth(sF))_{mn}\tilde\r^m\tilde\r^n +R(z,z')}
I(z,z')\,.
\la{K_-+1}
\ee

Finally, we have to push the operator ${\cal O}(s)$ in
(\ref{K_-+1}) on the right. This makes all objects $s$-dependent,
\be
K_{+-}(z,z'|s)=-\frac1{8(i\pi s)^{3/2}}\frac{sB}{\sinh(sB)}
e^{isG^2}
e^{\frac i4(F\coth(sF))_{mn}\tilde\r^m(s)\tilde\r^n(s) +R(z,z'|s)+\int_0^s dt \,\Sigma(t)}
I(z,z')\,,
\la{K_-+2}
\ee
where $R(z,z'|s)={\cal O}(s)R(z,z'){\cal O}(-s)$ and $\Sigma(t)$
is given in (\ref{Sigma(t)},\ref{Sigma}). The formula (\ref{K_-+2}) can be
identically rewritten as
\be
K_{+-}(z,z'|s)=-\frac1{8(i\pi s)^{3/2}}\frac{sB}{\sinh(sB)}
e^{isG^2}
e^{\frac i4(F\coth(sF))_{mn}\tilde\r^m(s)\tilde\r^n(s) +R(z,z')+\int_0^s dt(R'(t)+\Sigma(t))}
I(z,z')\,.
\la{K_-+2_}
\ee
The expression for $R'(t)$ can be found explicitly,
$R'(t)={\cal O}(t)[\bar W^\alpha \bar D_\alpha - W^\alpha D_\alpha ,R]{\cal
O}(-t)$ and then combined with (\ref{Sigma}),
\bea
\label{R+Sigma}
 R'(t)+\Sigma(t)&=&{\cal O}(t)\bigg\{
2i\bar\zeta W G +2i( \zeta\bar \zeta\, W\bar W- \zeta W \,
\bar\zeta \bar W)
\\&&
+i\bar\zeta^2[W^2 - \zeta^\alpha W^\beta  D_\alpha W_\beta]
-\frac12\bar\zeta^\beta  W^\alpha
[\tilde\rho_{\beta\gamma}\bar D^\gamma \bar W_\beta
- \tilde\rho_{\alpha\gamma}  D^\gamma W_\beta]
\bigg\}{\cal O}(-t)\,.
\nn
\eea
The form (\ref{K_-+2_}) of the heat kernel $K_{+-}$ is more useful
for loop computations than (\ref{K_-+2}) since at coincident superspace points the
function $R(z,z')$ vanishes, $R(z,z')|_{\zeta\to0} =0$,
and does not contribute.


\section{Low-energy effective action in ${\cal N}=2$ supersymmetric electrodynamics}

\subsection{Classical action and background field setup}
The classical action of the ${\cal N}=2$, $d=3$ supersymmetric
electrodynamics reads
\be
S_{\cN=2}=\f{1}{e^2}\int d^7 z\, G^2 - \int d^7 z\,\left( \bar Q_+ e^{2V} Q_+
+ \bar Q_- e^{-2V} Q_- \right)
- \left( m\int d^5z \,Q_+Q_-  + {\it c.c.}\right),
\label{action0}
\ee
where $Q_{\pm}$ are chiral superfields with opposite charges with
respect to the gauge superfield $V$. This action appears by virtue
of the dimensional reduction from the action of $\cN=1$,
$d=4$ electrodynamics \cite{book,GGRS}. In
principle, in three-dimensional space-time one could add the
Chern-Simons term $\int d^7 z\, VG$ which does not appear from the
$\cN=1$, $d=4$ SQED by
dimensional reduction. However, in the present work we restrict
ourself to studies of the low-energy effective action in
three-dimensional supersymmetric electrodynamics without the
Chern-Simons term. Note that the latter does not appear as a
result of the radiative corrections since the action
(\ref{action0}) is parity even \cite{AHISS,NS,Redlich1,Redlich2} (see also
\cite{Dunne} for a review).

We are interested in the part of the low-energy effective action
which depends on the gauge superfield only, $\Gamma=\Gamma[V]$,
while the chiral superfields $Q_\pm$ are integrated out.
For this problem the background field method in the $\cN=2$, $d=3$
superspace \cite{BPSreview} appears to be useful. We split the
gauge superfield $V$ into the background $V$ and quantum $v$ parts
\be
V\to V+e\,v\,.
\la{split}
\ee
Upon this splitting the Maxwell term in (\ref{action0}) changes as
\be
\frac1{e^2}\int d^7z\, G^2\to \frac1{e^2}\int d^7z\, G^2
+\frac ie\int d^7z\, v D^\alpha W_\alpha
+\frac18\int d^7z\, vD^\alpha \bar D^2 D_\alpha v\,.
\ee
The operator $D^\alpha \bar D^2 D_\alpha$ in the last term is
degenerate and requires gauge fixing. In particular, the
Fermi-Feynman gauge is implemented by the following gauge-fixing term
\be
S_{\rm gf}=-\frac1{16}\int d^7z\, v\{ D^2,\bar D^2\} v\,.
\label{Sgf}
\ee
Adding this term to (\ref{action0}) we get
\bea
\label{Squant}
S_{\rm quantum}&=&S_2+S_{\rm int}\,,\\
S_{2}&=&-\int d^7z\left(
v\square v +\bar{\cal Q}_+{\cal Q}_+ +\bar{\cal Q}_-{\cal Q}_-
\right)
- \left( m\int d^5z \,{\cal Q}_+{\cal Q}_-  + {\it c.c.}\right)\,,
\\
S_{\rm int}&=& -2\int d^7 z\left[ e\left( \bar{\cal Q}_+{\cal Q}_+ -  \bar{\cal Q}_-{\cal Q}_- \right)v
 +e^2\left( \bar{\cal Q}_+{\cal Q}_+ +\bar{\cal Q}_-{\cal Q}_- \right)v^2
 \right]+O(e^3)\,,
\la{Sint}
\eea
where ${\cal Q}_\pm$ and $\bar{\cal Q}_\pm$ are covariantly
(anti)chiral superfields with respect to the background gauge
superfield,
\be
\bar{\cal Q}_+ = \bar Q_+ e^{2V}\,,\quad
{\cal Q}_+ = Q_+\,,\quad
\bar{\cal Q}_- = \bar Q_- e^{-2V}\,,\quad
{\cal Q}_- = Q_-\,.
\label{cov-Q}
\ee

The action $S_{\rm int}$ specifies the interaction vertices while
$S_2$ is responsible for the propagators,
\bea
i\langle{\cal Q}_+(z){\cal Q}_-(z')\rangle&=& -m G_+(z,z')\,, \nn \\
i\langle \bar{\cal Q}_+(z) \bar{\cal Q}_-(z')\rangle &=& m G_-(z',z)\,, \nn \\
i\langle {\cal Q}_+(z)\bar {\cal Q}_+(z')\rangle&=& G_{+-}(z,z')=G_{-+}(z',z)\,,\nn\\
i\langle \bar{\cal Q}_-(z) {\cal Q}_-(z')\rangle&=&G_{-+}(z,z')\,,
\label{propagators}
\eea
where the Green's functions $G_+$ and $G_{+-}$ are defined by
the equations (\ref{2.43}) and (\ref{G+-def}), respectively.
The propagator for the real superfield $v$ reads
\be
2i\langle v(z)\,v(z')\rangle =G_0(z,z')\,,
\label{v-prop}
\ee
where
\be
G_0(z,z')= i\int_0^\infty ds\, K_0(z,z'|s)\,e^{-s\en}\,, \qquad
K_0(z,z'|s) = \f1{(4i\pi s)^{3/2}}
e^{\frac{i\rho^m \rho_m}{4s}} \zeta^2 \bar\zeta^2\,.
\la{photon1}
 \ee
Here $\rho^m$, $\zeta^\alpha$ and $\bar\zeta^\alpha$ are the
components of supersymmetric interval (\ref{superinterval}).

\subsection{Loop expansion and general structure of the effective
action} 
\begin{figure}[t]
\begin{center}
\setlength{\unitlength}{1mm}
\begin{picture}(150,50)
\thicklines
\qbezier(35,25)(35,30.74)(30.6,35.6)
\qbezier(30.6,35.6)(25.75,40)(20,40)
\qbezier(20,40)(14.26,40)(9.39,35.6)
\qbezier(9.39,35.6)(5,30.74)(5,25)
\qbezier(5,25)(5,19.26)(9.39,14.39)
\qbezier(9.39,14.39)(14.26,10)(20,10)
\qbezier(20,10)(25.74,10)(30.6,14.39)
\qbezier(30.6,14.39)(35,19.26)(35,25)
\qbezier(75,25)(75,30.74)(70.6,35.6)
\qbezier(70.6,35.6)(65.75,40)(60,40)
\qbezier(60,40)(54.26,40)(49.39,35.6)
\qbezier(49.39,35.6)(45,30.74)(45,25)
\qbezier(45,25)(45,19.26)(49.39,14.39)
\qbezier(49.39,14.39)(54.26,10)(60,10)
\qbezier(60,10)(65.74,10)(70.6,14.39)
\qbezier(70.6,14.39)(75,19.26)(75,25)
\qbezier(135,25)(135,30.74)(130.6,35.6)
\qbezier(130.6,35.6)(125.75,40)(120,40)
\qbezier(120,40)(114.26,40)(109.39,35.6)
\qbezier(109.39,35.6)(105,30.74)(105,25)
\qbezier(105,25)(105,19.26)(109.39,14.39)
\qbezier(109.39,14.39)(114.26,10)(120,10)
\qbezier(120,10)(125.74,10)(130.6,14.39)
\qbezier(130.6,14.39)(135,19.26)(135,25)
\put(5,25){\circle*{2}}
\put(35,25){\circle*{2}}
\put(45,25){\circle*{2}}
\put(75,25){\circle*{2}}
\put(105,25){\circle*{2}}
\put(135,25){\circle*{2}}
\put(39,24){$+$}
\put(2,35){$Q_+$}
\put(32,35){$\bar Q_+$}
\put(42,35){$Q_-$}
\put(72,35){$\bar Q_-$}
\put(102,35){$Q_+$}
\put(132,35){$Q_-$}
\put(2,13){$\bar Q_+$}
\put(32,13){$Q_+$}
\put(42,13){$\bar Q_-$}
\put(72,13){$Q_-$}
\put(102,13){$\bar Q_+$}
\put(132,13){$\bar Q_-$}
\put(8,27){$v$}
\put(30,27){$v$}
\put(48,27){$v$}
\put(70,27){$v$}
\put(108,27){$v$}
\put(130,27){$v$}
\qbezier(5,25)(6,27)(7,25)
\qbezier(7,25)(8,23)(9,25)
\qbezier(9,25)(10,27)(11,25)
\qbezier(11,25)(12,23)(13,25)
\qbezier(13,25)(14,27)(15,25)
\qbezier(15,25)(16,23)(17,25)
\qbezier(17,25)(18,27)(19,25)
\qbezier(19,25)(20,23)(21,25)
\qbezier(21,25)(22,27)(23,25)
\qbezier(23,25)(24,23)(25,25)
\qbezier(25,25)(26,27)(27,25)
\qbezier(27,25)(28,23)(29,25)
\qbezier(29,25)(30,27)(31,25)
\qbezier(31,25)(32,23)(33,25)
\qbezier(33,25)(34,27)(35,25)
\qbezier(45,25)(46,27)(47,25)
\qbezier(47,25)(48,23)(49,25)
\qbezier(49,25)(50,27)(51,25)
\qbezier(51,25)(52,23)(53,25)
\qbezier(53,25)(54,27)(55,25)
\qbezier(55,25)(56,23)(57,25)
\qbezier(57,25)(58,27)(59,25)
\qbezier(59,25)(60,23)(61,25)
\qbezier(61,25)(62,27)(63,25)
\qbezier(63,25)(64,23)(65,25)
\qbezier(65,25)(66,27)(67,25)
\qbezier(67,25)(68,23)(69,25)
\qbezier(69,25)(70,27)(71,25)
\qbezier(71,25)(72,23)(73,25)
\qbezier(73,25)(74,27)(75,25)
\qbezier(105,25)(106,27)(107,25)
\qbezier(107,25)(108,23)(109,25)
\qbezier(109,25)(110,27)(111,25)
\qbezier(111,25)(112,23)(113,25)
\qbezier(113,25)(114,27)(115,25)
\qbezier(115,25)(116,23)(117,25)
\qbezier(117,25)(118,27)(119,25)
\qbezier(119,25)(120,23)(121,25)
\qbezier(121,25)(122,27)(123,25)
\qbezier(123,25)(124,23)(125,25)
\qbezier(125,25)(126,27)(127,25)
\qbezier(127,25)(128,23)(129,25)
\qbezier(129,25)(130,27)(131,25)
\qbezier(131,25)(132,23)(133,25)
\qbezier(133,25)(134,27)(135,25)
\put(34,3){Type A}
\put(114,3){Type B}
  \end{picture}
\end{center}
\caption[b]{Two-loop supergraphs in $\cN=2$ supersymmetric electrodynamics.}
\label{fig1}
\end{figure}
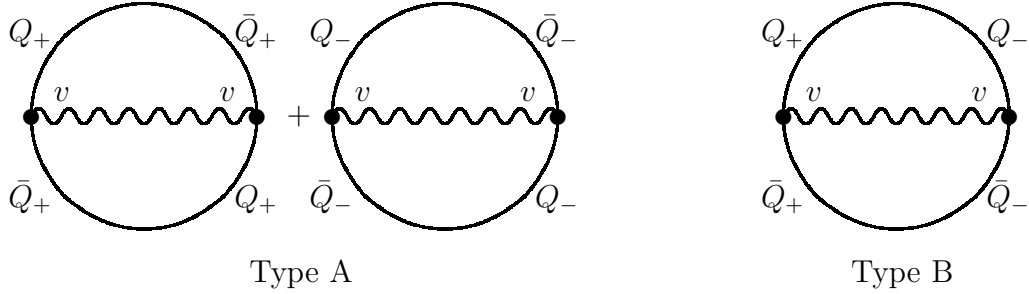

Within the present considerations we restrict ourself to the
two-loop effective action in the $\cN=2$ supersymmetric
electrodynamics (\ref{action0}),
\bea
\Gamma_{\cN=2}&=&\Gamma_{\cN=2}^{(1)}+\Gamma_{\cN=2}^{(2)}\,,\\
\Gamma^{(1)}_{\cN=2}&=&i\Tr \ln (\square_+ + m^2)\,,\\
\Gamma^{(2)}_{\cN=2}&=&-2e^2\int d^7z\, d^7z'[
G_{+-}(z,z')G_{+-}(z',z) +m^2 G_+ (z,z')G_-(z,z')]G_0(z,z')\,.
\label{loop-exp}
\eea
Here $\Gamma^{(1)}_{\cN=2}$ and $\Gamma^{(2)}_{\cN=2}$ are the
one- and two-loop contributions, respectively. The covariant
d'Alembertian $\square_+$ is given in (\ref{D_+}) while the Green's
functions $G_{+-}$, $G_+$ and $G_0$ are expressed through the heat
kernels as in (\ref{K_-+2}), (\ref{K+fin}) and (\ref{photon1}). The
two-loop effective action $\Gamma_{\cN=2}^{(2)}$ is represented by the
Feynman graphs in Fig.\ \ref{fig1}. The supergraphs of Types A
and B correspond to the two terms in the r.h.s.\ of (\ref{loop-exp}).
In principle, there could be a two-loop graph of topology `eight', but it vanishes since
the super-photon propagator (\ref{photon1}) is equal to zero at
coincident superspace points.

The one-loop effective action in three-dimensional $\cN=2$ SQED was
calculated in \cite{BPS1} (see also \cite{BPSreview}),
\bea
\Gamma^{(1)}_{\cN=2}&=&\frac1{2\pi}\int d^7z\Big[ G \ln
(G+\sqrt{G^2+m^2}) -\sqrt{G^2+m^2}\Big]\nn\\&& +\frac1{8\pi} \int d^7z
\int_0^\infty \frac{ds}{\sqrt{i\pi s}}e^{is(G^2+m^2)} \frac{W^2 \bar
W^2}{B^2}\left( \frac{\tanh(sB/2)}{sB/2}-1 \right) ,
\label{Gamma(1)} \eea
where $B^2$ is defined in (\ref{B2}), $B^2=
\frac12 D^2 W^2$. Recall that we consider the constant superfield
background (\ref{constr2}) subject to the supersymmetric Maxwell
equation (\ref{constr1}). Hence, the low-energy effective action is
a functional which depends on the superfield strengths $G$,
$W_\alpha$, $\bar W_\alpha$ and only on the first Grassmann
derivative of $W_\alpha$, i.e., on $N_{\alpha\beta}=D_{(\alpha}
W_{\beta)}$. All higher derivatives of the superfield strengths
either vanish or reduce to functions of $N_{\alpha\beta}$. As a
consequence, the two-loop effective action have the following general
superfield structure:
\be
 \Gamma^{(2)}_{\cN=2}=\f{e^2}{16\pi^3}\int d^7z\left[{\cal L}_1(G,B)
 + W^\a{\cal L}_{2\,\a}{}^\b(G,N) \bar W_\b + {\cal L}_3 (G,B) W^2\bar W^2
   \right]\,.
   \la{pre}
\ee
Here ${\cal L}_1$, ${\cal L}_{2\,\alpha}{}^\beta$ and ${\cal L}_3$
are some functions of $G$ and $N_{\alpha\beta}$ to be found from
direct quantum computations. Note that the contributions of the
form ${\cal L}_1$ and ${\cal L}_2$ are impossible in the
four-dimensional supersymmetric gauge theory. We will show that
these terms do appear in the two-loop effective action in the
three-dimensional supersymmetric electrodynamics.

\subsection{Computing two-loop diagrams}
\label{sec-2loop}
The two-loop diagrams of Type A in Fig.\ \ref{fig1}
correspond to the following contribution to the effective action:
\be
\Gamma_{\rm A}=-2e^2\int d^7z\, d^7z'\,
G_{+-}(z,z')G_{+-}(z',z)G_0(z,z')\,.
\la{A}
\ee
The propagators $G_{+-}$ and $G_0$ are expressed in terms of the
heat kernels as in (\ref{G+-K}) and (\ref{photon1}). Hence,
\bea
\Gamma_{\rm A}&=&
2i e^2 \int d^7z\, d^7z' \int_0^\infty ds\,dt\,du\,
 K_0(z,z'|u) K_{+-}(z,z'|s)K_{+-}(z',z|t)\, e^{i\,m^2(s+t)}\nn\\
&=&2i e^2 \int d^7z\, d^3\rho
 \int_0^\infty \frac{ds\,dt\,du}{(4i\pi u)^{3/2}}
e^{i\frac{\rho^m \rho_m }{4u}}e^{i\, m^2(s+t)}
K_{+-}(z,z'|s)K_{+-}(z',z|t)\bigg|_{\zeta\to0}.
\label{3.18}
\eea
Here we have taken into account that the heat kernel $K_0$ given
by (\ref{photon1}) contains $\zeta^2\bar\zeta^2$ which is
nothing but the delta-function over the Grassmann variables.
Hence, the expression (\ref{3.18}) involves the integration over
only one set of Grassmann variables, but we need to evaluate the heat kernel
$K_{+-}$ at coincident points, $\theta=\theta'$.
Evaluating this limit is a very straightforward, but tedious
procedure. Some details of this procedure are collected in
Appendix B. Here we present the result:\footnote{Each of the two heat kernels
$K_{+-}$ in (\ref{3.18}) contains the parallel displacement propagator
$I(z,z')$. These propagators cancel each other owing to the identity
(\ref{id-unit}) and the rest of (\ref{3.18}) depends only on superfield
strengths and their derivatives. Therefore we omit $I(z,z')$ further
and do not write it explicitly in the heat kernel.}
 \bea
K_{+-}(z,z'|s)\bigg|_{\z\rightarrow0}&=&-\f{1}{8(i\pi s)^{3/2}}\f{sB}
{\sinh(sB)}\,e^{isG^2}
\exp\bigg\{\f{i}{4}(F\coth(sF))_{mn}\,\r^m\r^n \nn\\
&&+iG W^\a f_{\a}{}^{\b}(s) \bar W_\b+W^\a \r_m f^m_{\a\b}(s) \bar W^\b
+\f{i}{2} W^2\bar W^2 f(s)
\bigg\}\,,\la{kerlim}
 \eea
where the following functions have been introduced:
 \bea
f_{\a}{}^{\b}(s)&=&2 B^{-2}(1-sN - e^{-sN})_{\a}{}^{\b}\,, \nn\\
f(s)&=&\f{1}{sB^4} \bigg[(sB)^2
-4\sinh^2(sB/2)\big(1+sB\tanh(sB/2)\big)\bigg]\,,\nn\\
f^m_{\a\b}(s)&=&\f12 B^{-2}(\cosh(sB)-1) \bigg[
(e^{-sN})_\b{}^\g N_\a{}^\d\,(\g^m)_{\g\d}+
(N(e^{-sN}))_\b{}^\d\,(\g^m)_{\a\d} \bigg]-    \la{fun}\\
& -& \f{1}{2}(F\coth(sF))^m{}_n\g^n_{\g\d}\bigg[
\big(\f{e^{-s N}-1}{N}\big)_\a{}^\g\, \big(\f{e^{- s N}-1}{N}\big)_\b{}^\d
+\frac{\ep_{\a\b}N^{\g\d}}{B^{3}}(sB-\sinh(sB)) \bigg]\,. \nn
 \eea
Note that (\ref{3.18}) includes also $K_{+-}(z',z|s)$ which has
the same form as (\ref{kerlim}), but the superspace points should
be swapped, $z \leftrightarrow z'$, or $\rho_m \to -\rho_m$. Hence, substituting
(\ref{kerlim}) into (\ref{3.18}) we find
 \bea
\Gamma_{\rm A}
&=&\frac{4ie^2}{(4i\pi)^{9/2}}\int d^7 z d^3\rho\int_0^\infty
\frac{ds\, dt\, du}{(s t u)^{3/2}}
e^{i(s+t)(G^2+m^2)}\f{st B^2}{\sinh (sB) \sinh (tB)}
e^{\f i2W^2 \bar W^2(f(s)+f(t))}
\nn\\
&&\times\exp\left[\frac i4\r A \r
+\r_m W^\a(f^m_{\a\b}(s)-f^m_{\a\b}(t))\bar W^\b
-2i G W^\a(f_\a{}^\b(s)+f_\a{}^\b(t))\bar W_\b
\right],
\la{gI2}
 \eea
where
 \be
A_{mn}(s,t,u)=(F \coth (sF))_{mn} + (F\coth (tF))_{mn} +\frac{\eta_{mn}}u
\label{180}
 \ee
is a symmetric $3\times3$ matrix with Lorentz indices.
It is convenient to express this matrix in terms of Lorentz
projectors $P^+$ and $P^-$,
 \be
A_{mn}=P^+_{mn}(a+u^{-1})+P^-_{mn}(b+u^{-1})\,,
\label{Aproj}
 \ee
where
\be
 a(s,t)=B\coth (sB)+ B\coth (tB) \,, \qquad
b(s,t)=s^{-1}+t^{-1}
\label{ab}
\ee
and
 \be
P^+_{mn}=\eta_{mn}+\frac1{4B^2}(N_{\alpha\beta}\gamma^{\alpha\beta}_m)
(N_{\gamma\delta}\gamma^{\gamma\delta}_n)\,,\qquad
P^-_{mn}=-\frac1{4B^2} (N_{\alpha\beta}\gamma^{\alpha\beta}_m)
(N_{\gamma\delta}\gamma^{\gamma\delta}_n)\,.
\label{projectors}
 \ee
These matrices obey standard properties of projection operators,
\be
(P^+)^2 = P^+\,, \quad
(P^-)^2 = P^-\,, \quad
P^+ P^- =0\,,\quad P^+_{mn} + P^-_{mn} = \eta_{mn}\,.
\ee

The integration over $d^3\rho$ in (\ref{gI2}) is simply Gaussian,
\be
\int d^3\rho\, e^{\frac i4 \rho_m A^{mn}\rho_n+\rho_m W^\a(f^m_{\a\b}(s)-f^m_{\a\b}(t))\bar W^\b}
=-\frac{(4\pi i)^{3/2}}{\sqrt{\det A}}e^{\f{i}2 W^2\bar W^2 {\cal F}(s,t,u)}\,,
\la{Gauss}
\ee
where
\be
{\cal F}(s,t,u)=- \f12 \big(f^m_{\a\b}(s)-f^m_{\a\b}(t)\big)(A^{-1})_m{}^n
\big(f_n^{\a\b}(s)-f_n^{\a\b}(t)\big)\,. \la{Fcudr}
\ee
Owing to the representation of the matrix $A_{mn}$ in terms of projectors
(\ref{Aproj}), it is easy to find its
determinant and the inverse,
\be
\frac1{\sqrt{\det A}}=\f1{(a +u^{-1})(b +u^{-1})^{1/2}}\,,
\qquad
(A^{-1})_{mn}=\frac{P^+_{mn}}{a+u^{-1}}
+\frac{P^-_{mn}}{b+u^{-1}}\,.
\la{a-1}
\ee
Now, using the explicit expression for the function $f^m_{\alpha\beta}$
(\ref{fun}) and the projectors $P^\pm$ (\ref{projectors}) we
compte the contractions of all indices in (\ref{Fcudr}),
 \be
{\cal F}(s,t,u)=\frac{F^+(s,t)}{a+u^{-1}}+\frac{F^-(s,t)}{b+u^{-1}}\,,
 \ee
where
 \bea
F^-(s,t)&=&\f1{B^6}\left[\left(\f{\sinh (sB)}{s} -\f{\sinh (tB)}{t} \right)^2
-\left(\f{\cosh (sB) -1}{s}-\f{\cosh (tB) -1}{t}\right)^2\right]\,,\nn\\
F^+(s,t)&=&-\f2{B^2}\bigg[(\cosh (2 B s)+2)\tanh^2\left(\f{Bs}{2}\right)
-2(\cosh(B(s-t))+\cosh(2 B(s-t))\nn\\
&+&\cosh (B (s+t)) \tanh\left(\f{B t}{2}\right)\tanh \left(\f{B s}{2}\right)
   +(\cosh (2 B t)+2) \tanh^2\left(\f{B t}{2}\right)\bigg]\,.
   \nn\\
 \eea

As the final step, we expand the exponent in the second line of \eq{gI2}
in a series\footnote{Actually, the series terminates owing to the
Grassmann nature of superfield strengths, $W_\alpha W_\beta W_\gamma\equiv 0$.}
and compute the integrals over $du$:
\bea
&&\int_0^\infty\frac{du}{u^{3/2}(a+u^{-1})(b+u^{-1})^{1/2}}
=\frac{2\,{\rm arccosh}\sqrt{a/b}}{\sqrt{a(a-b)}}\,,
\qquad (a>b) \label{int-du} \\
&&\int_0^\infty\frac{du}{u^{3/2}(a+u^{-1})(b+u^{-1})^{1/2}}\left[
\frac{F^+}{a+u^{-1}}+\frac{F^-}{b+u^{-1}}
\right]\nn\\
&=& \frac1{a-b}\left(
\frac{2F^-}{b}-\frac{F^+}{a}\right)+
\frac{2a(F^+-F^-)-bF^+}{(a(a-b))^{3/2}}\,{\rm arccosh}\sqrt{a/b}\,.
\label{int-du2}
\eea
As a result, the contribution to the low-energy
effective action from the diagrams of Type A in Fig.\ \ref{fig1}
matches previously discussed superfield structure (\ref{pre}),
 \be
\Gamma_{\rm A}=\frac{e^2}{16\pi^3}\int d^7z\left(
{\cal L}_1+W^\alpha {\cal L}_{2\alpha}{}^\beta \bar W_\beta
+W^2 \bar W^2 {\cal L}^{({\rm A})}_3
\right)\,,\la{GI}
 \ee
with ${\cal L}_1$, ${\cal L}_2$ and ${\cal L}_3$ given by
\bea
\label{L1}
{\cal L}_1&=&\int_0^\infty\frac{ds\, dt}{\sqrt{st}}e^{i(s+t)(G^2+m^2)}
     \frac{ B^2}{\sinh (sB) \sinh (tB)}
     \frac{2\,{\rm arccosh}\sqrt{a/b}}{\sqrt{a(a-b)}}\,,\\
\label{L2}
{\cal L}_{2\alpha}{}^\beta
&=& -\frac{2iG}{B^2}\int_0^\infty\frac{ds\,dt}{\sqrt{st}} e^{i(s+t)(G^2+m^2)}
    \frac{B^2}{\sinh (sB) \sinh (tB)}
    \frac{2\,{\rm arccosh}\sqrt{a/b}}{\sqrt{a(a-b)}}\nn\\
&&  \times(e^{-sN}-1+sN + e^{-tN}-1 +tN)_\alpha{}^\beta\,,\\
{\cal L}^{({\rm A})}_3
&=& \frac i2\int_0^\infty\frac{ds\,dt}{\sqrt{st}}e^{i(s+t)(G^2+m^2)}
    \frac{B^2}{\sinh (sB) \sinh (tB)}
    \frac1{a-b}\left(\frac{2F^-}{b} - \frac{F^+}{a} \right)\nn\\
&+& \int_0^\infty\frac{ds\,dt}{\sqrt{st}}e^{i(s+t)(G^2+m^2)}
    \frac{B^2}{\sinh sB \sinh tB}\bigg[
    \frac i2(f(s)+f(t)+\frac{a(F^+-F^-)-b F^+/2}{a(a-b)})
    \nn\\
&-& \frac{G^2}{B^4}((sB-\sinh sB + tB -\sinh tB)^2
    -(\cosh sB +\cosh tB -2)^2)
\bigg]\frac{2\,{\rm arccosh}\sqrt{a/b}}{\sqrt{a(a-b)}}\,. \nn\\
    \label{L3}
\eea

Consider now the diagram of Type B in Fig.\
\ref{fig1},
\bea
\Gamma_{\rm B}&=&-2e^2 m^2\int d^7z\, d^7z'\, G_+ (z,z')G_-(z,z')G_0(z,z')
\label{3.38}\\
&=&2i e^2 m^2  \int d^7z\, d^7z' \int_0^\infty
ds\, dt\, du\, K_+ (z,z'|s)K_-(z,z'|t)K_0(z,z'|u)
e^{im^2(s+t)}\,.\nn
\eea
Recall that the heat kernel $K_0$, given by (\ref{photon1}),
contains $\zeta^2\bar \zeta^2$ which is nothing but
the delta-function over the Grassmann variables. Hence,
the integration over only one set of Grassmann variables remains,
\be
\Gamma_{\rm B}=2ie^2m^2\int d^7z\, d^3 \rho \int_0^\infty
\frac{ds\,dt\,du}{(4i\pi u)^{3/2}} e^{i\frac{\rho^m \rho_m}{4u}}
e^{im^2(s+t)}
K_{+}(z,z'|s)K_{-}(z',z|t)\big|_{\zeta\to0}\,.
\label{Ga}
\ee
The heat kernel $K_+$ is given explicitly by (\ref{K+fin}).
Owing to the identity
\be
\zeta^2(s)|_{\zeta\to0}=s^2 W^2
\frac{\sinh^2(\frac{sB}2)}{(\frac{sB}2)^2}\,,\la{zs}
\ee
it is easy to find the limit $\zeta\to 0$ in
(\ref{K+fin}),\footnote{Here we omit the expressions for parallel
displacement propagators since they cancel in (\ref{Ga}) due to
(\ref{id-unit}).}
 \be
K_{+}(z,z'|s)\Big|_{\z\to0}=\frac{1}{4(i\pi s)^{3/2}}\frac{s
W^2}B \tanh \frac{sB}{2} e^{is G^2} e^{\frac i4(F \coth (sF))_{mn}\r^m \r^n} \,.
\label{K+limit}
\ee
Indeed, the expression (\ref{zs}) contains $W^2$ and this prevents
appearing of other superfield contributions which could come from the exponent
in (\ref{K+fin}). The antichiral heat kernel $K_{-}(z,z'|s)$ in
the limit $\zeta\to0$ has the same structure as (\ref{K+limit}),
but one should replace $W^2\rightarrow \bar W^2$.

Substituting (\ref{K+limit}) into (\ref{Ga}), we get
\be
\Gamma_{\rm B}=\frac{i e^2 m^2 }{64(i\pi)^{9/2}} \int d^7z\frac{W^2 \bar W^2}{B^2}
\int_0^\infty\frac{ds\, dt\, du}{\sqrt{st}u^{3/2}}
e^{i(s+t)(G^2+m^2)}\tanh\frac{sB}{2}\tanh\frac{tB}{2}
\int d^3\rho\,e^{\frac i4 \rho^m A_{mn}\rho^n}\,,
\label{3.42}
\ee
where the matrix $A_{mn}$ is given by (\ref{180}).
The Gaussian integral over $d^3\rho$ in (\ref{3.42}) is computed
according to (\ref{Gauss}),
\be
\int d^3\rho\, e^{\frac i4 \rho_m A^{mn}\rho_n}
=-\frac{(4\pi i)^{3/2}}{(a +u^{-1})(b +u^{-1})^{1/2}}\,.
\ee
The integral over $du$ is evaluated in (\ref{int-du}). As a result,
we find the contribution to the low-energy effective action from
the diagram of Type B in Fig.\ \ref{fig1} in the form
\be
\Gamma_{\rm B}=\frac{e^2}{16\pi^3}\int d^7z\,
W^2 \bar W^2 {\cal L}^{(B)}_3\,, \la{GII}
 \ee
with
\be
{\cal L}^{({\rm B})}_3 =\frac{4 m^2}{B^2}
\int_0^\infty\frac{ds\,dt}{\sqrt{st}}e^{i(s+t)(G^2+m^2)}
\tanh\frac{sB}{2}\tanh\frac{tB}{2}\frac{{\rm arccosh}\sqrt{a/b}}{\sqrt{a(a-b)}}\,.
\label{L4}
\ee

To summarize, the two-loop effective
action in the $\cN=2$ supersymmetric electrodynamics is given by
\be
 \Gamma^{(2)}_{\cN=2}=\Gamma_{\rm A} + \Gamma_{\rm B}=\f{e^2}{16\pi^3}\int d^7z\left[{\cal L}_1
 + W^\a{\cal L}_{2\,\a}{}^\b \bar W_\b + ({\cal L}^{({\rm A})}_3
  +{\cal L}^{({\rm B})}_3) W^2\bar W^2
   \right]\,,
 \label{3.46}
\ee
where the functions ${\cal L}_1$, ${\cal L}_2$, ${\cal L}_3^{({\rm A})}$
and ${\cal L}_3^{({\rm B})}$ are given by (\ref{L1}), (\ref{L2}),
(\ref{L3}) and (\ref{L4}), respectively.
It is important to note that all these functions are free of UV
quantum divergences because the integrations over
$s$ and $t$ are regular. This
is not surprising since the three-dimensional electrodynamic
without the Chern-Simons term is superrenormalizable because the gauge
coupling is dimensionful, $[e^2]=1$. Quantum divergences may
appear in the sector of effective K\"{a}hler potential, but the
Euler-Heisenberg effective action for the gauge superfield is
UV-finite.

\label{sec}

\subsection{Two-loop moduli space metric}
A lot of information about low-energy dynamics of supersymmetric
gauge theories is encoded in the structure of moduli space.
The analysis of moduli spaces in three-dimensional supersymmetric gauge
theories plays important role in studying the aspects of mirror symmetry
\cite{IS96,deBoer1996,deBoer1996ck,deBoer97,deBoer,AHISS} and
Seiberg-like dualities \cite{deBoer,AHISS,Aharony97,Karch,GK08} (see also
\cite{IS13,2} for very recent discussions of these problems). The
perturbative quantum corrections to the moduli space metric
in the $\cN=2$, $d=3$ gauge theories are known only up the
one-loop order \cite{IS96,deBoer1996,deBoer}. In the present
section we derive two-loop quantum corrections to this metric
which are stipulated by the effective action (\ref{3.46}).

The moduli space in $\cN=2$, $d=3$ supersymmetric
gauge theories is a K\"{a}hler manifold which is two-dimensional in
our case. It can be parametrized by two real coordinates $r$ and
$\sigma$. The coordinate $r$ is naturally identified with the vev
of the scalar field $\phi$ which is a part of the $\cN=2$, $d=3$
gauge multiplet, $r=\langle \phi \rangle$. This scalar is the
lowest component of the superfield strength $G$,
\be
G|_{\theta\to0}=\phi\,.
\ee
Another scalar field $a$ appears upon dualizing the Abelian
vector $A_m$,
\be
\partial_m a \propto \varepsilon_{mnp}F^{np}\,,
\ee
 where $F_{mn}$ is the Maxwell field strength corresponding to the Abelian vector
field $A_{m}$.\footnote{Here we use the same notation for the
usual Maxwell field strength as for the superfield $F_{mn}$
introduced in (\ref{gauge-alg}). The former appears as the lowest
component of the latter. We hope that this does not lead to any
confusions.} The coordinate $\sigma$ corresponds to the
vev of this scalar, $\sigma=\langle a \rangle$. In the present
section we find the metric on the moduli space parametrized by $r$
and $\sigma$, \be ds^2 = g_{rr}(r,\sigma) dr^2 +
g_{\sigma\sigma}(r,\sigma) d\sigma^2\,. \label{metric} \ee

The procedure of deriving the metric (\ref{metric}) from the
low-energy effective action is well described in \cite{deBoer}. The
moduli space metric is defined by the part of low-energy effective
action which is given by the full superspace Lagrangian of the
superfield strength $G$ without derivatives,
\be
S_{\rm low-energy}=\int d^7z \, f(G)\,.
\label{Slow-energy}
\ee
The classical action (\ref{action0}) and the one-loop effective
action (\ref{Gamma(1)}) contribute to $f(G)$ as
\bea
f^{(0)}(G)&=& \frac1{e^2} G^2\,,\label{f0}\\
f^{(1)}(G)&=&\frac1{2\pi}\Big[
G \ln (G+\sqrt{G^2+m^2}) -\sqrt{G^2+m^2}\Big]\,.
\label{f1}
\eea

To obtain the two-loop contribution to $f(G)$ we need to
evaluate the limit $B\to 0$ in the part of the effective action
(\ref{L1}). Taking into account the explicit form of the functions
$a(s,t)$ and $b(s,t)$ given in (\ref{ab}) we find
\be
\lim_{B\to0}\frac{{\rm arccosh}\sqrt{a/b}}{\sqrt{a(a-b)}} =
\frac{s\,t}{s+t}\,.
\ee
Substituting this expression into (\ref{L1}) and computing the
integrals over the parameters $s$ and $t$ we get
\be
\lim_{B\to0}{\cal L}_1 = -2\pi \ln(G^2+m^2)\,.
\ee
Hence, the two-loop contribution to $f(G)$ reads
\be
f^{(2)}(G)=-\frac{e^2}{8\pi^2}\ln(G^2+m^2)\,.
\label{f2}
\ee
Summarizing (\ref{f0}), (\ref{f1}) and (\ref{f2}) we conclude
\be
f(G)=\frac1{e^2}G^2 + \frac1{2\pi}[G\ln( G +\sqrt{G^2+m^2})
 -\sqrt{G^2+m^2}
  -\frac{e^2}{4\pi}\ln( G^2+m^2)]\,.
\ee

Given the function $f(G)$ one dualizes the linear superfield $G$
into a chiral superfield $\Phi$ as is described in \cite{HKLR}.
The chiral superfield serves as
the Lagrange multiplier for the linearity constraint
(\ref{linearity}),
\be
S_{\rm low-energy}=\int d^7z\, [f(G)-G(\Phi+\bar\Phi)]\,.
\label{Sl-e}
\ee
The superfield $G$ is treated now as unconstrained. Varying
(\ref{Sl-e}) with respect to $G$ we get
\be
\Phi+\bar\Phi=f'(G)=\frac2{e^2}G + \frac1{2\pi}\ln(
G+\sqrt{G^2+m^2})
-\frac{e^2}{4\pi^2}\frac G{G^2+m^2}\,.
\label{Phi+Phi}
\ee
From this equation the superfield $G$ should be expressed in terms of
$\Phi+\bar\Phi$ and substituted back to (\ref{Sl-e}). This yields
a sigma-model action,
\be
S_{\rm low-energy}=\int d^7z \, K(\Phi+\bar\Phi)\,,
\ee
with some function $K(\Phi+\bar\Phi)$ which is hard to write down
explicitly. However, we do not need the manifest expression for
$K$ since the K\"{a}hler metric is defined rather by its second derivative,
\be
ds^2 = K''\, d\Phi\, d\bar\Phi\,.
\label{ds2}
\ee
This metric should be expressed in terms of $r$ and
$\sigma$ where $r=\langle G\rangle$ and $\sigma$ can be identified
with the imaginary part of $\Phi$, $\sigma={\rm Im}\Phi$. Using
the fact that the inverse Legendre transform is a Legendre
transform, we have
\be
K'(\Phi+\bar\Phi)=r\,.
\ee
From this equation and from (\ref{Phi+Phi}) we conclude
\be
K''(\Phi+\bar\Phi)=\left(\frac{\partial(\Phi+\bar\Phi)}{\partial r}
\right)^{-1} =\frac12 \frac1{g(r)}\,,
\label{K''}
\ee
where
\be
g(r)=\frac1{e^2}+\frac1{4\pi}\frac1{\sqrt{r^2+m^2}}
  + \frac{e^2}{8\pi^2} \frac{r^2-m^2}{(r^2+m^2)^2}\,.
\label{g(r)}
\ee

Finally, we note that (\ref{Phi+Phi}) implies that
\be
d\Phi= g(r) dr
 +i d\sigma\,,\qquad
d\bar\Phi= g(r)dr
 -i d\sigma\,.
\ee
Substituting now (\ref{K''}) and the latter identities into
(\ref{ds2}) we find the moduli space metric in the form
\be
ds^2=\frac12 g(r) dr^2+
\frac12\frac1{g(r)}d\sigma^2\,.
\ee
In the massless limit the function $g(r)$ in (\ref{g(r)})
simplifies such that
\be
ds^2|_{m=0}=\frac12\left(\frac1{e^2}+\frac1{4\pi r}
  + \frac{e^2}{8\pi^2 r^2}
\right)dr^2  +
\frac12\left(\frac1{e^2}+\frac1{4\pi r}
  + \frac{e^2}{8\pi^2 r^2}
\right)^{-1}d\sigma^2\,.
\label{moduli-metric}
\ee

Equation (\ref{moduli-metric}) shows that the one-loop metric is corrected
by the two-loop contribution $\frac{e^2}{8\pi^2 r^2}$. It is
naturally to expect that the $n$-loop correction could be of the
form $c_n \frac1{e^2}(\frac{e^2}{r})^n$, with some coefficient $c_n$. It
is very tempting to compute such higher-loop coefficients $c_n$
and to find a closed expression for all-loop moduli space metric
both for the Abelian and non-Abelian $\cN=2$, $d=3$ gauge theories.
In principle, it could resolve the singularity of the moduli space
metric at small $r$.\footnote{An alternative mechanism for
resolving the singularity of the moduli space metric was proposed
recently in \cite{STT}.}

\section{Low-energy effective action in $\cN=4$ supersymmetric
electrodynamics}
\subsection{Classical action and structure of two-loop effective action}
The classical action of the $\cN=4$ supersymmetric electrodynamics
reads
\bea
\label{N4}
S_{\cN=4}&=&\frac1{e^2} \int d^7z (G^2-\frac12 \bar\Phi\Phi)
+S_Q\,,\\
S_Q&=&-\int d^7z(\bar{\cal Q}_+ {\cal Q}_+ + \bar{\cal  Q}_- {\cal Q}_-)
-\int d^5z \,{\cal Q}_+ \Phi{\cal Q}_-
+\int d^5\bar z\, \bar{\cal Q}_+ \bar\Phi\bar{\cal Q}_-
\,,\nn
\eea
where the covariantly chiral superfields ${\cal Q}_\pm$ are related
to the standard chiral superfields $Q_\pm$ as in (\ref{cov-Q}).
The action (\ref{N4}) is invariant under the following `hidden'
$\cN=2$ supersymmetry,
\bea
\delta V&=&\frac12(\bar\epsilon^\alpha \bar\theta_\alpha \Phi
 - \epsilon^\alpha \theta_\alpha \bar \Phi)\,,\nn\\
\delta \Phi&=&i\epsilon^\alpha W_\alpha\,,\qquad
 \delta\bar \Phi= i\bar\epsilon^\alpha \bar W_\alpha\,,\nn\\
\delta{\cal Q}_+&=&-\frac14 \bar\nabla^2(\bar\epsilon^\alpha \bar\theta_\alpha \bar
{\cal Q}_-)\,,\qquad
\delta{\cal Q}_-=\frac14 \bar\nabla^2(\bar\epsilon^\alpha \bar\theta_\alpha \bar
{\cal Q}_+)\,,\nn\\
\delta \bar{\cal Q}_+&=&-\frac14\nabla^2(\epsilon^\alpha\theta_\alpha
{\cal Q}_-)\,,\qquad
\delta \bar{\cal Q}_-=\frac14\nabla^2(\epsilon^\alpha\theta_\alpha
{\cal Q}_+)\,,
\eea
where $\epsilon^\alpha$ and $\bar\epsilon^\alpha$ are the
supersymmetry parameters. Note that the action (\ref{N4}) appears from
the $\cN=2$, $d=4$ electrodynamics by means of the dimensional reduction.
Two-loop Euler-Heisenberg effective action in the
latter was studied in \cite{Kuz0310}.

The $\cN=4$ gauge multiplet is described by the pair $(V,\Phi)$.
We make the background-quantum splitting for both these
superfields,
\be
V\to V+e\,v \,,\qquad
\Phi\to \Phi+ e\, \phi\,,
\ee
while the hypermultiplet $({\cal Q}_+, {\cal Q}_-)$ is considered as
the `quantum' superfield which should be integrated out in the
path integral. The background gauge superfield $V$ obeys the
constraints (\ref{constr1}) and (\ref{constr2}) while $\Phi$ is
simply constant,
\be
D_\alpha \Phi=0\,.
\label{Phi-const}
\ee
Upon quantization in the Fermi-Feynman gauge (\ref{Sgf}), we end
up with the following action for `quantum' superfields,
\bea
S_{\rm quantum}&=&S_2 + S_{\rm int}\,,\\
S_2&=&-\int d^7z (v\square v+ \frac12 \bar\phi \phi
 +\bar{\cal Q}_+{\cal Q}_+ + \bar{\cal Q}_- {\cal Q}_-)
-\left(\int d^5z \, {\cal Q}_+ \Phi {\cal Q}_- +c.c.\right),\\
S_{\rm int}&=& -2\int d^7 z\left[ e\left( \bar{\cal Q}_+{\cal Q}_+ -  \bar{\cal Q}_-{\cal Q}_- \right)v
 +e^2\left( \bar{\cal Q}_+{\cal Q}_+ +\bar{\cal Q}_-{\cal Q}_- \right)v^2
 \right]\nn\\
&&-e\int d^5z \, {\cal Q}_+ \phi {\cal Q}_-
+e\int d^5 \bar z\, \bar{\cal Q}_+ \bar\phi \bar{\cal Q}_+ +
O(e^3)\,.
\label{4.7}
\eea

The propagators for the hypermultiplets and for the gauge superfield
$V$ are the same as in the $\cN=2$ electrodynamics,
(\ref{propagators}) and (\ref{v-prop}), but the mass parameter $m$
is now promoted to the background superfield $\Phi$.
Additionally, there is the propagator for the chiral superfield
$\phi$,
\be
\langle \phi(z)\bar\phi(z')\rangle = -\frac i8 \bar D^2 D^2
G_0(z,z')\,.
\label{chiral-prop}
\ee
There are also vertices with the chiral superfield
$\phi$ represented in the last line in (\ref{4.7}). Owing to these
propagators and vertices with the chiral superfield $\phi$ the
two-loop effective action in the $\cN=4$ electrodynamics gets
additional contribution $\Gamma_{\rm C}$ as compared with the $\cN=2$ case
(\ref{loop-exp}),
\bea
\label{N4loop-exp}
\Gamma^{(2)}_{\cN=4}&=& \Gamma_{\rm A}+\Gamma_{\rm B}+\Gamma_{\rm C}\,,\\
\Gamma_{\rm A}&=&-2e^2\int d^7z\, d^7z'\,
G_{+-}(z,z')G_{+-}(z',z)G_0(z,z')\,,\\
\Gamma_{\rm B}&=&-2e^2 \int d^7z\, d^7z'\, \bar\Phi \Phi\, G_+
(z,z')G_-(z,z')G_0(z,z')\,,\label{N4GB}\\
\Gamma_{\rm C}&=&2e^2\int d^7z\, d^7 z'\, G_{+-}(z,z')G_{+-}(z,z')G_0(z,z')\,.
\label{N4GC}
\eea

The part of the effective action $\Gamma_{\rm A}$ takes into account the
graphs of Type A in Fig.\ \ref{fig1} which
are exactly the same as in the $\cN=2$ electrodynamics.
Therefore we can borrow the result (\ref{GI}) for $\Gamma_{\rm A}$ from
the effective action of the $\cN=2$ electrodynamics.

The term $\Gamma_{\rm B}$ in (\ref{N4loop-exp}) corresponds to the
supergraph of Type B in Fig.\ \ref{fig1}. The expression
(\ref{N4GB}) has the same form as (\ref{3.38}), but the mass
parameter $m$ should now be replaced with the chiral superfield
$\Phi$. Since we consider the background with constant chiral
superfield (\ref{Phi-const}), the result of computing this diagram
is given by (\ref{GII}), where one should replace $m^2\to
\bar\Phi\Phi$.

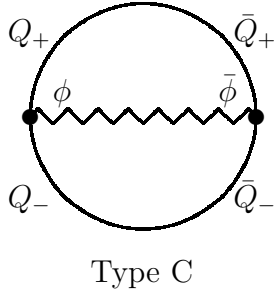
\begin{figure}[t]
\begin{center}
\setlength{\unitlength}{1mm}
\begin{picture}(50,50)
\thicklines
\qbezier(35,25)(35,30.74)(30.6,35.6)
\qbezier(30.6,35.6)(25.75,40)(20,40)
\qbezier(20,40)(14.26,40)(9.39,35.6)
\qbezier(9.39,35.6)(5,30.74)(5,25)
\qbezier(5,25)(5,19.26)(9.39,14.39)
\qbezier(9.39,14.39)(14.26,10)(20,10)
\qbezier(20,10)(25.74,10)(30.6,14.39)
\qbezier(30.6,14.39)(35,19.26)(35,25)
\put(5,25){\circle*{2}}
\put(35,25){\circle*{2}}
\put(2,35){$Q_+$}
\put(32,35){$\bar Q_+$}
\put(2,13){$Q_-$}
\put(32,13){$\bar Q_-$}
\put(8,27){$\phi$}
\put(30,27){$\bar\phi$}
\qbezier(5,25)(6,26)(6,26)
\qbezier(6,26)(7,25)(8,24)
\qbezier(8,24)(9,25)(10,26)
\qbezier(10,26)(11,25)(12,24)
\qbezier(12,24)(13,25)(14,26)
\qbezier(14,26)(15,25)(16,24)
\qbezier(16,24)(17,25)(18,26)
\qbezier(18,26)(19,25)(20,24)
\qbezier(20,24)(21,25)(22,26)
\qbezier(22,26)(23,25)(24,24)
\qbezier(24,24)(25,25)(26,26)
\qbezier(26,26)(27,25)(28,24)
\qbezier(28,24)(29,25)(30,26)
\qbezier(30,26)(31,25)(32,24)
\qbezier(32,24)(33,25)(34,26)
\qbezier(34,26)(35,25)(35,25)
\put(13,3){Type C}
  \end{picture}
\end{center}
\caption[b]{Two-loop supergraph in $\cN=4$ supersymmetric electrodynamics
which involves chiral superfield propagator $\langle \phi\bar\phi\rangle$.}
\label{fig2}
\end{figure}

The term $\Gamma_{\rm C}$ in (\ref{N4loop-exp}) is new as compared with the
$\cN=2$ supersymmetric electrodynamics since it involves the
propagator of the chiral superfield (\ref{chiral-prop}).
It is represented by the supergraph of Type C in Fig.\ \ref{fig2}.

The one-loop effective action in the $\cN=4$ electrodynamics was
computed in \cite{BPS1}. It has the same form as (\ref{Gamma(1)}), but
the mass parameter $m$ should now be promoted to the background
superfield $\Phi$,
\bea
\Gamma^{(1)}_{\cN=4}&=&\frac1{2\pi}\int d^7z\Big[
G \ln (G+\sqrt{G^2+\bar\Phi\Phi}) -\sqrt{G^2+\Phi\bar\Phi}\Big]\nn\\&&
+\frac1{8\pi} \int d^7z \int_0^\infty
\frac{ds}{\sqrt{i\pi s}}e^{is(G^2+\Phi\bar\Phi)}
\frac{W^2 \bar W^2}{B^2}\left(
\frac{\tanh(sB/2)}{sB/2}-1
\right)\,.
\eea
In what follows, we concentrate on computing two-loop
contributions to the effective action.

\subsection{Computing two-loop effective action}
Consider the diagram of Type C in Fig.\ \ref{fig2}. Its analytic
expression (\ref{N4GC}) is very similar to (\ref{A}), but the
arguments of one of the Green's function are swapped. Hence, the
algorithm of computing this graph is the same as in
Sect.\ \ref{sec-2loop}.

Using the super-photon propagator (\ref{photon1}) and the
definition of the heat kernel $K_{+-}$ (\ref{G+-K}) we get
\be
\Gamma_{\rm C}=-2i e^2 \int d^7z\, d^3\rho
 \int_0^\infty \frac{ds\,dt\,du}{(4i\pi u)^{3/2}}
e^{i\frac{\rho^m \rho_m }{4u}}e^{i\, \bar\Phi\Phi(s+t)}
K_{+-}(z,z'|s)K_{+-}(z,z'|t)\bigg|_{\zeta\to0}.
\label{4.14}
\ee
The propagator $K_{+-}$ at coincident Grassmann
points is given by (\ref{kerlim}). Substituting this expression
into (\ref{4.14}) we find
 \bea
\Gamma_{\rm C}
&=&-\frac{4ie^2}{(4i\pi)^{9/2}}\int d^7 z d^3\rho\int_0^\infty
\frac{ds\, dt\, du}{(s t u)^{3/2}}
e^{i(s+t)(G^2+\bar\Phi\Phi)}\f{st B^2}{\sinh (sB) \sinh (tB)}
e^{\f i2W^2 \bar W^2(f(s)+f(t))}
\nn\\
&&\times\exp\left[\frac i4\r A \r
+\r_m W^\a(f^m_{\a\b}(s)+f^m_{\a\b}(t))\bar W^\b
-2i G W^\a(f_\a{}^\b(s)+f_\a{}^\b(t))\bar W_\b
\right],
\label{GamC}
 \eea
where the matrix $A$ is given in (\ref{180}) and the functions $f$
are written down in (\ref{fun}). Note that (\ref{GamC}) differs
from (\ref{gI2}) only in one sign in the last line. Hence, many
cancellations occur among (\ref{GamC}) and (\ref{gI2}).

Let us compute the Gaussian integral over $d^3\rho$ in
(\ref{GamC}),
\be
\int d^3\rho\, e^{\frac i4 \rho_m A^{mn}\rho_n+\rho_m W^\a(f^m_{\a\b}(s)+f^m_{\a\b}(t))\bar W^\b}
=-\frac{(4\pi i)^{3/2}}{\sqrt{\det A}}e^{\f{i}2 W^2\bar W^2 \tilde{\cal F}(s,t,u)}\,,
\ee
where
\be
\tilde{\cal F}(s,t,u)=- \f12 \big(f^m_{\a\b}(s)+f^m_{\a\b}(t)\big)(A^{-1})_m{}^n
\big(f_n^{\a\b}(s)+f_n^{\a\b}(t)\big)\,. \la{tildeFcudr}
\ee
In the part of the effective action $\Gamma_{\rm A} + \Gamma_{\rm C}$ the
function (\ref{tildeFcudr}) appears in the following combination
with (\ref{Fcudr}),
\be
{\cal F}(s,t,u)-\tilde {\cal F}(s,t,u) =
2f^m_{\alpha\beta}(s)(A^{-1})_{mn} f^{m\, \alpha\beta}(t)\,.
\label{4.18}
\ee
Given the function $f_{\alpha\beta}^m$ in (\ref{fun}) and the
inverse matrix $A^{-1}$ in (\ref{a-1}) we compute the contractions
in (\ref{4.18}),
\be
{\cal F}(s,t,u)-\tilde {\cal F}(s,t,u) = \frac{{\cal F}^+(s,t)}{a+u^{-1}}
+\frac{{\cal F}^-(s,t)}{b+u^{-1}}\,,
\ee
where $a$ and $b$ are given in (\ref{ab}) and
\bea
{\cal F}^+(s,t)&=&\frac 8{B^2}\tanh\frac{sB}{2}\tanh\frac{tB}{2}
\left[
\cosh(B(s-t))+\cosh(2B(s-t))+\cosh(B(s+t))
\right],\nn\\
{\cal F}^-(s,t)&=&\frac4{s t B^4}
\left[
(\cosh sB -1)(\cosh tB-1)-
(\sinh sB -sB)(\sinh tB -tB)
\right].
\eea
Hence, for the sum of $\Gamma_{\rm A}$ and $\Gamma_{\rm C}$ we
find
\bea
\Gamma_{\rm A}+\Gamma_{\rm C}&=&\frac{i\,e^2}{32\pi^3}\int d^7z\,W^2 \bar W^2 \int_0^\infty
\frac{ds\,dt\,du}{(stu)^{3/2}}e^{i(s+t)(G^2+\bar\Phi\Phi)}\nn\\
&&\times\frac{stB^2}{\sinh s B\sinh tB}\frac1{(a+u^{-1})(b+u^{-1})}
\left[\frac{{\cal F}^+(s,t)}{a+u^{-1}}+\frac{{\cal F}^-(s,t)}{b+u^{-1}}
\right].
\eea
Finally, we perform the integration over $du$ using
(\ref{int-du2}),
\bea
\Gamma_{\rm A}+\Gamma_{\rm C}&=&\frac{i\,e^2}{32\pi^3}\int d^7z\,W^2 \bar W^2 \int_0^\infty
\frac{ds\,dt}{(st)^{3/2}}e^{i(s+t)(G^2+\bar\Phi\Phi)}\frac{stB^2}{\sinh s B\sinh tB}\nn\\
&&\times\left[
\frac1{a-b}\left(
\frac{2{\cal F}^-}{b}-\frac{{\cal F}^+}{a}\right)+
\frac{2a({\cal F}^+-{\cal F}^-)-b{\cal F}^+}{(a(a-b))^{3/2}}\,{\rm arccosh}\sqrt{\frac a b}
\right].
\label{GA+C}
\eea

The contribution from the diagram of Type B in Fig.\ \ref{fig1}
can be easily obtained from (\ref{GII},\ref{L4}),
\be
\Gamma_{\rm B}=\frac{e^2}{4\pi^3}\int d^7z\frac{W^2 \bar W^2 \Phi\bar\Phi}{B^2}
\int_0^\infty\frac{ds\,dt}{\sqrt{st}}e^{i(s+t)(G^2+\bar\Phi\Phi)}
\tanh\frac{sB}{2}\tanh\frac{tB}{2}\frac{{\rm
arccosh}\sqrt{a/b}}{\sqrt{a(a-b)}}\,.
\label{GB4}
\ee

We conclude that the two-loop low-energy effective action in the
$\cN=4$ SQED is given by (\ref{N4loop-exp}) with
$\Gamma_{\rm A}+\Gamma_{\rm C}$ and $\Gamma_{\rm B}$ written down explicitly in
(\ref{GA+C}) and (\ref{GB4}), respectively. As is seen from these
expressions, all contributions to the two-loop effective action
$\Gamma^{(2)}_{\cN=4}$ contain $W^2\bar W^2$ and the terms without
$W's$, like (\ref{L1}), do not appear. Hence, in the $\cN=4$ SQED
there are no two-loop quantum corrections to the low-energy
effective action of the form (\ref{Slow-energy}) and the moduli
space metric is one-loop exact.\footnote{This is the well-known Taub-NUT metric
derived in \cite{HKLR} from geometrical principles.} This is in
agreement with the conclusions of \cite{deBoer}.

\section{Summary and discussion}
We have developed a manifestly $\cN=2$ supersymmetric and
gauge-covariant technique for studying contributions to low-energy
effective actions in three-dimensional supersymmetric gauge theories
beyond one-loop order. As an application, we computed two-loop
Euler-Heisenberg effective actions in $\cN=2$ and $\cN=4$
supersymmetric electrodynamics in the $\cN=2$, $d=3$ superspace.

One of the features of the three-dimensional $\cN=2$ supersymmetric
gauge theory in comparison with the four-dimensional $\cN=1$ supersymmetric
gauge theory is that the gauge superfield has not
only `spinorial' superfield strengths $W_\alpha$ and $\bar
W_\alpha$, but also the `scalar' superfield strength $G$. As a
consequence, in the three-dimensional SQED, there are completely new
superfield contributions to the two-loop effective actions
stipulated by this superfield $G$ having no analogs in four
dimensions (see the two-loop effective action for $\cN=1, d=4$ gauge
theories in \cite{Kuz0310,Kuz07}). In particular, in the
$\cN=2$, $d=3$ SQED these new terms are found in the form (\ref{L1})
and (\ref{L2}). We argue that these expressions play important role
in the low-energy dynamics since they are responsible, in
particular, for the two-loop quantum corrections to the moduli space
metric. We explicitly computed the moduli space metric
(\ref{moduli-metric}) in the $\cN=2$ SQED which takes into account
the two-loop corrections. To the best of our knowledge, only
one-loop perturbatibe corrections to the moduli space have been
known so far \cite{deBoer,AHISS}.

Two-loop effective action in the $\cN=4$, $d=3$ SQED receives
additional contributions represented by the graph in Fig.\
\ref{fig2} as compared with the two-loop effective action of the
$\cN=2$ SQED described by the background field dependent Feynman
diagrams in Fig.\ \ref{fig1}. Indeed, the chiral superfield $\Phi$
becomes dynamical in the $\cN=4$ theory and propagates inside the
two-loop diagram. This leads to many cancellations among the
diagrams of Type A in Fig.\ 1 and Type C in Fig.\ \ref{fig2}. In
particular, the terms of the form (\ref{L1}) and (\ref{L2}) are
completely cancelled in the $\cN=4$ SQED two-loop effective action.
As a consequence, the moduli space in the $\cN=4$ electrodynamics is
not renormalized by two-loop corrections and remains one-loop exact.
This is in agreement with conclusions made in \cite{IS96,deBoer}.

Concerning technical details of two-loop computations
performed in the present work, we obtained
exact propagators of chiral superfields interacting with
slowly-varying background gauge superfield. The propagators
involve the so-called parallel displacement propagator $I(z,z')$
which provides the gauge covariance on all stages of quantum loop
computations. This technique is a three-dimensional
analog of the methods of covariant perturbative computations in
the $\cN=1$, $d=4$ superspace \cite{Kuz03}. We believe that the
properties of parallel displacement propagator $I(z,z')$ and the
exact propagators for chiral superfields in three-dimensional
gauge theories explored in the present paper will be useful for
studying low-energy effective actions in other three-dimensional
gauge theories including non-Abelian ones.

An important extension of the results of the present paper is the
inclusion of the Chern-Simons term into considerations. With the
non-trivial Chern-Simons term the form of super-photon propagator
(\ref{photon1}) changes \cite{Avdeev2} acquiring extra spinorial
derivatives on the full superspace delta-function. Careful
accounting of these derivatives in the two-loop computations
requires separate studies. However, the two-loop low-energy
effective actions in the three-dimensional supergauge models with
Chern-Simons terms are of high interest in the light of recent
discussions \cite{IS13,2}\footnote{One-loop effective action in
$\cN=2$ Chern-Simons gauge theory coupled to matter is studied in
\cite{BP}, \cite{BP1}.}. Next, it is important to study the
low-energy effective actions in the BLG \cite{BL1,BL2,BL3,BL4,G1,G2}
and ABJM \cite{ABJM} models which
should describe the low-energy dynamics of multiple M2 branes.

Another important extension of the present considerations is the
study of two-loop effective actions in non-Abelian three-dimensional
supersymmetric gauge theories. The one-loop effective action in
various three-dimensional super Yang-Mills models were found in
\cite{BPS2}, but the two-loop extension of these results remains
an open problem. Finally, it is tempting to study two-loop
quantum corrections to the K\" ahler potential in three-dimensional
$\cN=2$ gauge theories (the two-loop K\" ahler potential in the
$\cN=2$, $d=3$ sigma-models was obtained in \cite{BMS}.)

\vspace{3mm} {\bf Acknowledgements.} We are grateful to S. Kuzenko
for useful comments concerning the form of heat kernels in
four-dimensional supersymmetric electrodynamics, to N. Pletnev for
reading the manuscript and to D. Sorokin for stimulating
discussions. I.L.B.\ thanks the Galileo Galilei Institute for
Theoretical Physics for the hospitality and the INFN for partial
support during the completion of this work. B.S.M.\ is grateful to
INFN, Sezione di Padova and to ITP, University of Wroclaw for kind
hospitality and support. The authors are grateful to the RFBR
grant Nr.\ 12-02-00121 and LRSS grant Nr.\ 224.2012.2. The study
was partially supported by the Ministry of education and science
of Russian Federation, project 14.B.21.0774. The work of I.L.B.
and I.B.S. was also supported by the RFBR grants Nrs.\ 13-02-90430
and 13-02-91330 and DFG grant LE 838/12/1. B.S.M.\ acknowledges the support from RF Federal
Programs ``Kadry'' Nrs.\ 14.B37.21.1298 and 16.740.11.0469 and MSE Program ``Nauka''
Nr.\ 1.604.2011. The work of I.B.S.\ was supported by Marie
Curie research fellowship Nr.\ 909231, ``QuantumSupersymmetry''
and by the Padova University Project CPDA119349.


\appendix
\section{$\cN=2$ superspace conventions}\la{conventions}
\setcounter{equation}{0}
\renewcommand{\theequation}{A.\arabic{equation}}
In the present paper we use $\cN=2$, $d=3$ superspace conventions following previous
works \cite{BPS1,BPS2}. In particular, the gamma matrices
$(\gamma^0)_\alpha^\beta=-i\sigma_2$,
$(\gamma^1)_\alpha^\beta=\sigma_3$,
$(\gamma^2)_\alpha^\beta=\sigma_1$ obey the Clifford algebra
\be
\{ \gamma^m,\gamma^n\}=-2\eta^{mn}\,,\qquad
\eta^{mn}=\mbox{diag}(1,-1,-1)\,,
\ee
and the following orthogonality and completeness relations
\be
(\gamma^m)_{\alpha\beta}(\gamma^n)^{\alpha\beta}=2\eta^{mn}\,,\qquad
(\gamma^m)_{\alpha\beta}(\gamma_m)^{\rho\sigma}
=(\delta_\alpha^\rho\delta_\beta^\sigma+\delta_\alpha^\sigma\delta_\beta^\rho)\,.
\ee
We raise and lower the spinor indices with the $\varepsilon$-tensor, e.g.,
$(\gamma_m)_{\alpha\beta}=\varepsilon_{\alpha\sigma}(\gamma_m)^\sigma_\beta$,
$\varepsilon_{12}=1$.

Any vector index is converted into a
pair of spinor ones according to the following rules
\bea
&&x^{\alpha\beta}=(\gamma_m)^{\alpha\beta} x^m\,,\qquad
x^m=\frac12(\gamma^m)_{\alpha\beta}x^{\alpha\beta}\,,\nn\\
&&\partial_{\alpha\beta}=(\gamma^m)_{\alpha\beta}\partial_m\,,\qquad
\partial_m=\frac12(\gamma_m)^{\alpha\beta}\partial_{\alpha\beta}\,,
\eea
so that
\be
\partial_m x^n=\delta_m^n\,,\qquad
\partial_{\alpha\beta} x^{\rho\sigma}
=
\delta_\alpha^\rho\delta_\beta^\sigma+\delta_\alpha^\sigma\delta_\beta^\rho
=2\delta_\alpha^{(\rho}  \delta_\beta^{\sigma)}\,.
\ee

The covariant spinor derivatives
\be
D_\alpha=\frac\partial{\partial\theta^\alpha}+i\bar\theta^\beta
\partial_{\alpha\beta},\qquad
\bar D_\alpha=-\frac\partial{\partial\bar\theta^\alpha}
-i\theta^\beta \partial_{\alpha\beta}
\label{Dexpl}
\ee
obey the standard anticommutation relation
\be
\{D_\alpha, \bar D_\beta
\}=-2i\partial_{\alpha\beta}\,.
\label{Dalg}
\ee

The integration measure in the full $\cN=2$, $d=3$ superspace is
defined as
\be
d^7z\equiv d^3x d^4\theta=\frac1{16}d^3x\,D^2\bar D^2\,,\quad
\mbox{so that}\quad
\int d^3x\, f(x)=\int d^7z\,\theta^2\bar\theta^2 f(x)\,,
\label{fullmeasure}
\ee
for some field $f(x)$. Here we use the following conventions for
contractions of the spinor indices
\be
D^2= D^\alpha D_\alpha\,,\quad
\bar D^2=\bar D^\alpha\bar D_\alpha\,,\quad
\theta^2=\theta^\alpha\theta_\alpha\,,\quad
\bar\theta^2=\bar\theta^\alpha\bar\theta_\alpha\,.
\ee

The chiral subspace is parametrized by $z_+=(x_+^m,\theta_\alpha)$,
where $x_\pm^m=x^m\pm i\gamma^m_{\alpha\beta}\theta^\alpha\bar\theta^\beta$.
The integration measure in the chiral superspace
$d^5z\equiv d^3x d^2\theta$ is related to the
full superspace measure (\ref{fullmeasure}) as
\be
d^7z=-\frac14d^5z\,\bar D^2=-\frac14 d^5\bar z \,D^2\,.
\label{chiral-measure}
\ee

\section{Heat kernel $K_{+-}$ at coincident points}
\la{AppB}
\setcounter{equation}{0}
\renewcommand{\theequation}{B.\arabic{equation}}

Consider the heat kernel $K_{+-}$ given by the expression
(\ref{K_-+2}),\footnote{We omit the parallel displacement propagator
$I(z,z')$.}
\be
K_{+-}(z,z'|s)=-\frac1{8(i\pi s)^{3/2}}\frac{sB}{\sinh(sB)}
e^{isG^2}
e^X\,,
\ee
where
\be
X(s)={\frac i4(F\coth(sF))_{mn}\tilde\r^m(s)\tilde\r^n(s) +R(z,z')+\int_0^s
dt(R'(t)+\Sigma(t))}\,,
\ee
and $R'(t)+\Sigma(t)$ is given in (\ref{R+Sigma}),
\bea
 R'(t)+\Sigma(t)&=&
2i\bar\zeta^\alpha(t) W_\alpha(t) G +2i[
\zeta^\alpha(t)\bar \zeta_\alpha(t)\, W^\beta(t)\bar W_\beta(t)
- \zeta^\alpha(t) W_\alpha(t) \,
\bar\zeta^\beta(t) \bar W_\beta(t)]
\nn\\&&
+i\bar\zeta^2(t)W^2(t) -\frac12\bar\zeta^\beta(t)  W^\alpha(t)
[\rho_{\beta\gamma}(t)\bar D^\gamma \bar W_\beta
- \rho_{\alpha\gamma}(t)  D^\gamma W_\beta]\,.
\label{B3}
\eea
The $t$-dependent objects in the r.h.s.\ of (\ref{B3}) are given in
(\ref{id's}). Note that here we use the bosonic interval $\rho_{\alpha\beta}$
rather than its chiral version $\tilde\rho_{\alpha\beta}$ given in
(\ref{chirho}).
It is clear that the problem of computing the heat kernel $K_{+-}$
at coincident points is reduced to finding $X\Big|_{\zeta\to0}$.

First of all, we point out that the function $R(z,z')$ given in
(\ref{R}) vanishes in this limit,
\be
R(z,z')\Big|_{\zeta\to 0}=0\,.
\ee
Hence, we need to compute
\be
X(s)\Big|_{\zeta\to0}
=\left(\frac i4(F\coth(sF))_{mn}\tilde\r^m(s)\tilde\r^n(s) +\int_0^s
dt(R'(t)+\Sigma(t))\right)\bigg|_{\zeta\to 0}\,.
\label{B.5}
\ee

Consider $\tilde \rho^m(s)$. Using (\ref{chirho}) and
(\ref{id's}), it can be rewritten as
\be
\tilde\rho^m(s)=\tilde\rho^m-2i \gamma^m_{\alpha\beta}
 \int_0^s W^\alpha(t)\bar\zeta^\beta(t)dt\,.
\ee
We substitute here the expressions (\ref{id's}) for
$W^\alpha(s)$ and $\bar\zeta^\beta(s)$ and compute the integral
over $dt$,
\be
\tilde\rho^m(s)|=\rho^m+i(\gamma^m_{\alpha\beta}N^{\alpha\beta})
\frac{ W^\gamma \bar W_\gamma }{B^3}(sB-\sinh sB)
+i\gamma^m_{\alpha\beta}W^\gamma \bar W^\delta
\left(\frac{e^{-sN}-1}{N} \right)_\gamma{}^\alpha
\left(\frac{e^{-sN}-1}{N} \right)_\delta{}^\beta\,.
\ee
Here the following identities have been used
\be
(N^{2n})_\alpha^\beta = \delta_\alpha^\beta B^{2n}\,,\quad
(N^{2n+1})_\alpha^\beta = N_\alpha^\beta B^{2n}\,,\quad
B^2\equiv\frac12 N_\alpha^\beta N_\beta^\alpha\,.
\ee
Hence, for the first term in r.h.s.\ in (\ref{B.5}) we find
\bea
&&\frac i4(F\coth(sF))_{mn}\tilde\r^m(s)\tilde\r^n(s)|=
\frac i4(F\coth(sF))_{mn}\r^m\r^n\nn\\&
-&\frac12(F\coth(sF))_{mn}\r^m \gamma^n_{\alpha\beta}
\bigg(N^{\alpha\beta}
\frac{ W \bar W}{B^3}(sB-\sinh sB)
\nn\\&&
+W^\gamma \bar W^\delta
\left(\frac{e^{-sN}-1}{N} \right)_\gamma{}^\alpha
\left(\frac{e^{-sN}-1}{N} \right)_\delta{}^\beta\bigg)
\nn\\&+&
\frac i2\frac{W^2 \bar W^2 }{B^4}\left[
(2B\coth sB + \frac1s)(\cosh sB -1)^2
-\frac1s(\sinh sB-sB)^2 \right].
\label{B9}
\eea
In deriving this expression the following identities could be
useful
\be
(F\coth sF)_m{}^m = 2B\coth sB +\frac1s\,,\qquad
(F\coth sF)_{mn}(N\gamma^m)(N\gamma^n)=-4\frac{B^2}{s}\,.
\ee

Consider now the last term in (\ref{B.5}) which is given by $\int_0^s
dt(R'(t)+\Sigma(t))$. For this purpose we compute the limit of
coincident Grassmann points of various terms in (\ref{B3}),
\be
W^\alpha(s) W_\alpha(s)|=W^2\,,\quad
W^\alpha(s) \bar W_\alpha(s)|=W^\alpha \bar W_\alpha\,, \la{Ws}
\ee
\be
\zeta^2(s)|=\frac{4 W^2}{B^2}
\sinh^2(\frac{sB}2)\,,\qquad
\bar\zeta^2(s)|=\frac{4 \bar W^2}{B^2}
\sinh^2(\frac{sB}2)\,,
\ee
\bea
\zeta^\alpha(s)\bar W_\alpha(s)|&=&-\bar W^\alpha W_\alpha
\frac{\sinh(sB)}{B}
+\frac{W^\alpha \bar W^\beta N_{\alpha\beta}}{B^2}
 2\sinh^2\frac{sB}{2}\,,\\
\bar\zeta^\alpha(s) W_\alpha(s)|&=&-\bar W^\alpha W_\alpha
\frac{\sinh(sB)}{B}
-\frac{W^\alpha \bar W^\beta N_{\alpha\beta}}{B^2}
 2\sinh^2\frac{sB}{2}\,,\\
\zeta^\alpha(s)W_\alpha(s)|&=&-W^2 \frac{\sinh(sB)}{B}\,,\qquad
\bar\zeta^\alpha(s)\bar W_\alpha(s)|=-\bar W^2 \frac{\sinh(sB)}{B}\,,\\
(\zeta \bar W)(\bar \zeta W)|&=&-\frac{W^2 \bar W^2 }{B^2}
(\sinh^2 sB- \cosh sB +1)\,,\\
\zeta \bar\zeta\, W \bar W - \zeta W\, \bar\zeta \bar W|
&=&-2 \zeta W \, \bar \zeta \bar W - (\zeta \bar W )(\bar \zeta
W)\nn\\&
=&\frac{W^2 \bar W^2}{B^2}(1-\cosh sB -\sinh^2 sB)
\,,
\eea
\be
2i( \zeta\bar \zeta\, W\bar W- \zeta W \,
\bar\zeta \bar W)|+i\zeta^2 \bar W^2|=
-2i\frac{W^2 \bar W^2}{B^2}\sinh^2 sB\,.
\ee
Substituting these expressions into (\ref{B3}) and integrating
over the parameter $t$ we obtain
\bea
\label{B20}
\int_0^s dt (R'(t)+\Sigma(t))|&=&
\frac{2iG}B^2 W^\alpha \left(
e^{-sN}-1+sN\right)_\alpha{}^\beta \bar W_\beta
\\&&
-i\frac{W^2 \bar W^2}{B^3}(\sinh sB \cosh sB -sB)
\nn\\&&
-\frac12\rho^m (\cosh sB-1)\bar W^\alpha(e^{-sN})_\alpha{}^\beta
((\gamma_m)_\beta^\gamma N_\gamma^\delta - N_\beta^\gamma
(\gamma_m)_\gamma^\delta)W_\delta\,.\nn
\eea

Putting (\ref{B9}) and (\ref{B20}) together, we find
\be
X(s)\Big|=\frac i4(F \coth(sF))_{mn}\r^m\r^n
+\r_m f^m_{\alpha\beta}(s)W^\alpha \bar W^\beta +
\frac i2 W^2 \bar W^2f(s)
-i G W^\alpha f_\alpha{}^\beta(s)\bar W_\beta\,,
\ee
where
\bea
f_{\a}{}^{\b}(s)&=&2 B^{-2}(1-sN - e^{-sN})_{\a}{}^{\b}\,, \\
f(s)&=&\f{1}{sB^4} \bigg[(sB)^2
-4\sinh^2(sB/2)\big(1+sB\tanh(sB/2)\big)\bigg]\,,\nn\\
f^m_{\a\b}(s)&=&\f12 B^{-2}(\cosh(sB)-1) \bigg[
(e^{-sN})_\b{}^\g N_\a{}^\d\,(\g^m)_{\g\d}+
(N(e^{-sN}))_\b{}^\d\,(\g^m)_{\a\d} \bigg]-    \nn\\
& -& \f{1}{2}(F\coth(sF))_{n m}\g^m_{\g\d}\bigg[
\big(\f{e^{-s N}-1}{N}\big)_\a{}^\g\, \big(\f{e^{- s N}-1}{N}\big)_\b{}^\d
+\frac{\ep_{\a\b}N^{\g\d}}{B^{3}}(sB-\sinh(sB)) \bigg]\,. \nn
\eea



\end{document}